\documentclass[aps,prd,tighteness,preprint,superscriptaddress,showpacs,nofootinbib]{revtex4}
\usepackage{epsf,epsfig,graphics,graphicx}
\usepackage{verbatim,color,ulem}
\newcommand{\be}{\begin{equation}}
\newcommand{\ee}{\end{equation}}
\newcommand{\bea}{\begin{eqnarray}}
\newcommand{\eea}{\end{eqnarray}}
\newcommand{\ba}{\begin{array}}
\newcommand{\ea}{\end{array}}

\usepackage{amssymb,amsmath,amsthm}
\usepackage[mathscr]{eucal}

\usepackage[pdfencoding=auto,psdextra]{hyperref}
\usepackage{enumitem}
\newcommand{\RNum}[1]{\uppercase\expandafter{\romannumeral #1\relax}}
\usepackage[toc,page]{appendix}
\usepackage{cancel}
\usepackage{xcolor}
\newcommand*\Eval[3]{\left.#1\right\rvert_{#2}^{#3}}
\usepackage[hang, flushmargin]{footmisc}
\usepackage{footnotebackref}

\begin{document}

\title{Holographic approach to thermalization in general anisotropic theories}
\author{Po-Chun Sun}
\email{ 410414215@gms.ndhu.edu.tw} \affiliation{Department of Physics,
National Dong-Hwa University, Hualien, Taiwan, R.O.C.}
\author{Da-Shin Lee}
\email{dslee@mail.ndhu.edu.tw} \affiliation{Department of Physics,
National Dong-Hwa University, Hualien, Taiwan, R.O.C.}
\author{Chen-Pin Yeh}
\email{chenpinyeh@mail.ndhu.edu.tw} \affiliation{Department of
Physics, National Dong-Hwa University, Hualien, Taiwan, R.O.C.}

\begin{abstract}
We employ the holographic approach to study the
thermalization in the quenched strongly-coupled field theories
with very general anisotropic scalings including Lifshitz and
hyperscaling violating fixed points. The holographic dual
is a Vaidya-like time-dependent geometry where the asymptotic
metric has general anisotropic scaling isometries. We find the
Ryu-Takanayagi extremal surface and use it to calculate the time-dependent
entanglement entropy between a strip region with width $2R$ and
its outside region. In the special case with an isotropic metric, we also
explore the entanglement entropy for a spherical region of
radius $R$. The growth of the entanglement entropy
characterizes the thermalization rate after a quench. We study the thermalization process
in the early times and late times in both large $R$ and small $R$ limits. The allowed scaling parameter regions are constrained by the null energy conditions as well as the condition for the existence of the Ryu-Takanayagi extremal surfaces.
This generalizes the previous works on this subject.
All obtained results can be compared with experiments and other methods of probing thermalization.

\end{abstract}


\maketitle
\tableofcontents
\section{Introduction}
The holographic duality provides a unique method to investigate the dynamics of strongly coupled field theories, which links geometrical quantities to quantum observables. The uses of this interplay to study thermal phases of strongly coupled field theories by their holographically dual AdS black hole/brane geometries have attracted lots of attention since it was first porposed \cite{witten}. It was then suggested that the small perturbations of the AdS metric are dual to the hydrodynamics or linear response theories of the boundary CFTs \cite{Hydrodynamics,son_1,son_2}. Also, by probed strings in the AdS background, the diffusion behavior of Brownian particles in the boundary fields can be derived (see \cite{Brownian}for reviews). In particular, the time evolution of entanglement entropy between Brownian particles and boundary fields had been studied by means of the probed string method \cite{Yeh_19_1}.

Along this line of thoughts, it is very interesting to consider the holographic analysis of strongly coupled field theories far from equilibrium. For example, a system, which starts from
a highly excited state after a quench, is expected to evolve toward a stationary state at the thermal equilibrium. A holographic description of this far from equilibrium problem is a process of gravitational collapse ending in the formation of a black hole/brane, which in the simplest case can be described by a time-dependent Vaidya geometry.
According to Ryu and Takayanagi \cite{Ryu}, the entanglement entropy between a spatial region, $\Sigma$ of dimension $d-1$, and its outside region in the $d$-dimensional boundary theory is dual to the area of the extremal surface $\Gamma$ in the $d+1$-dimensional bulk geometry, which is holomorphic to $\Sigma$ and has the same boundary $\partial\Sigma$ as $\Sigma$.
The prescription for time-dependent holographic backgrounds was proposed in \cite{Hubeny_07}. The uses of this prescription to study the thermalization process following the quenches in various Vaidya-like backgrounds have been found in \cite{Abajo,Balasubramanian 1,Balasubramanian 2,Liu-s,Liu-d,Fonda,Galante,Farsam,Fondathesis,Irina,Curtis,Yong,Pawe,Ville,Pallab,Wu,Veronika,Gouteraux,Zhuang,Nozaki,Tameem,Alishahiha,eva,Ageev,Ecker,Cartwright19,Cartwright20,Mozaffar}.
In these studies, the region $\Sigma$ in the boundary theories is bounded
by $\partial\Sigma$, which is either a $d-2$-dimensional sphere of radius $R$ or two planes separated by a distance $2R$.
Computing the time-dependent entanglement entropy as a function of spatial
scales thus provides a probe of scale-dependent thermalization.
The bulk Vaidya geometry describes the falling of a $d-1$-dimensional thin shell along a light cone from the boundary at the time $t=0$. In the case of the boundary of a plane, a black brane eventually forms when the thin shell falls within the horizon distance, $y_h$. In the boundary theory, this process is dual to the input of energy at $t=0$ (quench) driving the system to a far from equilibrium state that subsequently thermalizes.

In this paper, we extend some of these works by considering the Vaidya-like geometries that describe the formation of black branes with the general asymptotic anisotropic scaling symmetries including Lifshitz and hyperscaling violation. A holographic model in the static anisotropic background has been used to study the diffusion of heavy quarks in the anisotropy plasma when they are slightly out of equilibrium \cite{Giataganas:2013zaa}, and also the dissipation and fluctuation of Brownian particles within the linear response regions \cite{Yeh_18_2}.  Here, in the cases of far-from-equilibrium states from holographic time-dependent anisotropic backgrounds, we consider the time evolution of entanglement entropy between a strip region of width $2R$ and its outside region. To have the analytical expressions we focus on both the large $R$ and small $R$ limits as compared to the horizon scale. In both cases, the early time and late time entanglement growth and its dependence on the scaling parameters are explored. Notice that in this study, the anisotropic effects are encoded in the effective bulk spacial dimension, $\tilde{d}$ as will be defined later, which is just $d$ in an isotropic theory. In addition, the entangled region we propose is to probe the anisotropic spatial coordinate with the scaling parameter, say $a_1$ relative to the scaling parameter $a_y$ in the bulk direction where there exists a free parameter $\Delta=a_1+a_y-4$ that is zero in \cite{Liu-s} and \cite{Fonda}.  Apart from the strip case, we study the entanglement region bounded by a sphere of radius $R$ when the backgrounds are isotropic but with $\Delta\neq 0$. We thus mainly focus on the contributions from the nonzero $\Delta$  to the dynamics. Additionally, the constraints on these scaling parameters from the null energy conditions together with the constraints from the existence of the solution of the extremal surface are derived.  Then the obtained holographic entanglement entropy can be used to justify (or falsify) the holographic method from the  experiment tests and other methods on strongly coupled problems.

The layout of the paper is as follows. In Sec. \ref{sec1} we introduce a Lifshitz-like anisotropic hyperscaling violation theory, and the corresponding static black hole metric. The null energy conditions are developed to impose the constraints on the scaling parameters of the theory. The dynamics of the extremal surface and its equations of motion are studied in Sec. \ref{sec2}. In Sec. \ref{sec3}, we first compute the entanglement entropy at the thermal equilibrium. The early time entanglement entropy growth and  late time saturation will be computed later and then discussed in Sec. \ref{sec4} and \ref{sec5} respectively. In Sec. \ref{sec6}, we choose the Einstein-Axion-Dilaton theory as an example to realize the allowed scaling parameter regions given by the constraints from the null energy conditions as well as the condition for the existence of the extremal surfaces. Finally, Sec. \ref{sec7} concludes the work. In Appendix \ref{app}, we provide the detailed analysis about the regions of the scaling parameters for the system to have either continuous or discontinuous saturation in the strip case.

\section{The holography background} \label{sec1}
To make our analysis of thermalization as general as possible, we
consider the gravitational collapse that eventually forms the following black brane with the metric
  \be \label{metric}
  ds^2=g_{\mu\nu}dx^{\mu}dx^{\nu}=-y^{-a_0}h(y)d^2t+\frac1{y^{4-a_y}h(y)}d^2y+\sum_{i=1}^{d-1}y^{-a_i}d^2x_i \, ,
  \ee
where near the boundary $y=0$, the blackening factor is assumed to be $h(0)= 1$. Then, in the boundary, the metric has scaling symmetries,
 \be
 \label{scaling}
 y \rightarrow \lambda^{-1} y,~~t \rightarrow \lambda^{1-\frac{a_y}{2}-\frac{a_0}{2}} t,~~x_i \rightarrow \lambda^{1-\frac{a_y}{2}-\frac{a_i}{2}}
 x_i,~~g_{\mu\nu}\rightarrow \lambda^{a_y-2} g_{\mu\nu}\,.
 \ee
We also assume that there is a simple zero for $h(y)$ at $y=y_h$,
corresponding to the position of the horizon. Thus, near the boundary,
$h(y)$ has the leading term,
  \be
  \label{BHb}
  h(y)=1-M y^{{\Delta_h}}
  \ee
with
\begin{equation}\label{bk_c}
{\Delta_h}\geq 1\, .
\end{equation}
The temperature of the black brane is given by $T=\frac{|h'(y_h)|}{4\pi
y_h^{\frac{a_0}{2}+\frac{a_y}{2}-2}}$. In order to describe the
formation of the black brane of (\ref{metric}), we introduce the
Eddington-Finkelstein coordinates given by
   \be\label{efcoor}
   v=t-\int_0^y\frac{y'^{\frac{a_y}{2}+\frac{a_0}{2}-2}}{h(y')}dy' \, .
   \ee
Then the metric in (\ref{metric}) becomes
   \be
   \label{BHef}
   ds^2=-y^{-a_0}h(y)d^2 v-2
   y^{\frac{a_y}{2}-\frac{a_0}{2}-2}dvdy+\sum_{i=1}^{d-1}y^{a_i}dx_i^2 \, .
   \ee
In this work, we consider the gravitational collapse with the following Vaidya-type metric
   \be
   \label{vaidya}
   ds^2=-y^{-a_0}f(v,y)d^2 v-2
   y^{\frac{a_y}{2}-\frac{a_0}{2}-2}dvdy+\sum_{i=1}^{d-1}y^{a_i}dx_i^2 \, ,
   \ee
where $f(v,y)=1-\Theta(v)g(y)$, and $g(y)=1-h(y)$. The metric describes
the formation of the black brane of (\ref{metric}) by  the infalling
of a $d-1$-dimensional delta-function shell along the trajectory
$v=0$. The region $v>0$  outside the shell has the metric as in
(\ref{metric}) whereas the region $v<0$  inside the shell has the pure hyperscaling violating anisotropic Lifshitz metric of the form  (\ref{metric}) by setting $h=1$.
From the viewpoint of the boundary theory,
this gives a quench on the system at $t=0$ and subsequently the system evolves into a thermal equilibrium state.

Nevertheless, the null energy conditions (NECs) constrain the parameters in the metric \cite{Dimitrio_19}, obtained as
\begin{equation}\label{null}
T_{\mu \nu }\ell^{\mu } \ell^{\nu } \geq 0\qquad \mbox{for}~~  \ell_{\mu } \ell^{\mu }=0\, .
\end{equation}
In the Einstein gravity, NECs are equivalent to $R_{\mu \nu }\ell^{\mu } \ell^{\nu } \geq 0$ where $R_{\mu \nu}$ is the curvature tensor obtained from (\ref{vaidya}), and then give the following constraint equations,
    \begin{align}
\sum_{j=1}^{d-1}\Big[a_j\left( 2 a_0-\Delta _0-a_j-2\right)\Big] & \geq 0 \, , \label{nec1}\\
a_1(\tilde{d}-1) g(y) \partial_v \Theta(v) \frac{y^{\frac{\Delta_0}{2} +1}}{f(v,y)}+\left(a_0-a_i\right)\left(\Delta_0+2+a_1(\tilde{d}-1) \right)f(v,y)&\nonumber\\
+{\Delta_h} g(y) \Theta (v) \left(2(a_0-a_i)+\Delta_0+2-2{\Delta_h}+a_1(\tilde{d}-1) \right)& \geq 0 \, ,   ~~~i=1,2,3...d-1 \, , \label{nec2} \\
4a_1(\tilde{d}-1) g(y) \partial_v\Theta(v)y^{\frac{\Delta_0}{2} +1}+f(v,y)^2\sum_{j=1}^{d-1}\Big[a_j\left( 2 a_0-\Delta _0-a_j-2 \right)\Big]& \geq 0 \label{nec3}
\end{align}
with $\tilde{d}=\sum_{i=1}^{d-1}\frac{a_i}{a_1}+1$, and $\Delta_0=a_0+a_y-4$.  In the case of an isotropic background with all $a_i$'s to be equal,  NECs reduce to the ones in the hyperscaling violating Lifshitz theory ~\cite{Fonda}.

\section{Dynamics of the extremal surface} \label{sec2}

In this section, we derive the equations of motion for the extremal surfaces when the entanglement region $\Sigma$ in the boundary theory is either a strip of width $2R$ in $x_1$ direction or a region bounded by a sphere of radius $R$. In this work, we consider the $d+1$-dimensional bulk geometry. Thus the boundary theory living at $y=0$ is in a $d$-dimensional spacetime. The entanglement region $\Sigma$ in the boundary theory is bounded by a $d-2$-dimensional surface $\partial\Sigma$. We consider that the infalling planar shell propagates from the boundary at the boundary time $t=0$. Starting from this section, the time $t$ means the boundary time and is different from the coordinate time $t$ in the previous section.
Then the $d-1$-dimensional extremal surface $\Gamma$ in the bulk, if exists, is uniquely fixed by the boundary conditions where the extremal surface should match $\partial\Sigma$ at the boundary $y=0$ and at the boundary time $t$. Once we find the time-dependent extremal surface $\Gamma$ and its area $A_{\Gamma}$, the entanglement entropy between $\Sigma$ and its outside region is given by the Ryu-Takayanagi formula
\begin{equation} \label{S_A}
S(R,t)=\frac{A_{\Gamma}}{4}.
\end{equation}
The approach we adopt in this paper mainly follow the works of \cite{Liu-s} and \cite{Fonda}. Here
we highlight  key equations and show relevant solutions in the following sections, which are straightforward generalizations of their results to our model.
\subsection{sphere}

We first consider  $\partial\Sigma$ as a sphere with radius $R$.
Thus, to accommodate the $d-2$-dimensional rotational symmetry, we assume the metric (\ref{vaidya}) with such a symmetry
by setting all $a_i$'s to be euqal, say $a_i=a_1$. In the spherical coordinates, the metric in the following form
   \be
   \label{sphere}
  ds^2=-y^{-a_1+\Delta -\Delta _0}f(v,y)d^2 v-2
   y^{-a_1+\Delta -\frac{\Delta _0}{2}}dvdy+y^{a_1}\left(d^2\rho+\rho^2d^2\Omega_{d-2}\right)
   \ee
 where $\Delta=a_1+a_y-4$. The embedding of the $d-1$-dimensional extremal surface
 $\Gamma$ in (\ref{sphere}) can be described by two functions
 $v(\rho)$ and $y(\rho)$ together with an extension in all $\Omega_{d-2}$
 directions. The area of $\Gamma$ is then given by
   \be
   \label{area}
  A_{\Gamma}=\frac{A_{\partial \Sigma}}{R^{d-2}}\int_0^{R}d\rho\frac{\rho^{d-2}}{y^{\frac{a_1}{2}(d-1)}}\sqrt{Q}\, ,
   \ee
where
   \be
   \label{Q}
 Q=\big\vert 1-y^{\Delta -\Delta _0}\dot{v}^2 f(y,v)-2 y^{\Delta -\frac{\Delta _0}{2}}\dot{v} \,
   \dot{y}\big\vert
   \ee
 with $\dot v=\frac{d}{d\rho}v(\rho)$ and $\dot
 y=\frac{d}{d\rho}y(\rho)$. $A_{\partial \Sigma}=\frac{2\pi^{\frac{d-1}{2}}}{\Gamma(\frac{d-1}{2})}R^{d-2}$ is the area of a $d-2$-sphere with radius $R$. The functions $v(\rho)$ and $y(\rho)$
 are determined by minimizing the area in (\ref{area}),
 giving the  equations of motion,
   \begin{equation}\label{eqmot1}
   \frac{\sqrt{Q} y^{\Delta_0-\Delta+\frac{a_1 (d-1)}{2}}}{\rho ^{d-2}} \frac{d}{d\rho}\left(\frac{\rho ^{d-2} y^{-\frac{a_1 (d-1)}{2}-\Delta_0+\Delta}}{\sqrt{Q}} \left(\dot{v} f(y,v)+y^{\frac{\Delta_0}{2}} \dot{y}\right)\right)=\frac{1}{2}\dot{v}^2 \frac{\partial f(y,v)}{\partial v} \, ,
\end{equation}
\begin{align}\begin{split}\label{eqmot2}
    &\frac{\sqrt{Q} y^{\frac{a_1 (d-1)}{2}+\Delta_0-\Delta}}{\rho ^{d-2}} \frac{d}{d\rho}\left(\frac{\rho ^{d-2} y^{\frac{1}{2} \left(-a_1 (d-1)+2 \Delta -\Delta _0\right)}}{\sqrt{Q_\rho}} \dot{v}\right)\\
    &=\frac{1}{2}\left(a_1 (d-1) Q y^{\Delta_0-\Delta-1}-\left(\Delta _0-2 \Delta\right) y^{\frac{1}{2} \left(\Delta_0-2\right)} \dot{v} \dot{y}+\frac{\dot{v}^2}{y} \left(\left(\Delta-\Delta_0\right) f(y,v)+y \frac{\partial f(y,v)}{\partial y}\right)\right)\, .
\end{split}\end{align}
The boundary conditions are
  \be
  \label{bcyv}
  \dot v(0)=\dot y(0)=0,~~v(R)=t,~~y(R)=0\, .
  \ee
Again, here and later, the time $t$ will label the boundary time.

Since $f(y,v)=1$ for $v < 0$ and $f(y,v)=h(y)$ for $v>0$,  $\frac{\partial f(y,v)}{\partial v}=0$ in both the $v<0$ and $v>0$ regions.
From  (\ref{eqmot1}) there exists a constant of motion $E$,
\begin{equation}\label{es}
\frac{\rho ^{d-2}}{\sqrt{Q} y^{\frac{a_1}{2}(d-1)}}
\left(\frac{\dot{v}
f(y,v)}{y^{\Delta_0-\Delta}}+\frac{\dot{y}}{y^{\frac{1}{2} \left(\Delta _0-2 \Delta \right)}}\right)\equiv
E={\rm constant} \,.
  \end{equation}
Solving (\ref{es}) for $\dot v$ gives
  \be
  \label{vdot}
  \frac{\dot v}{y^{\frac{a_0}{2}+\frac{a_y}{2}-2}}=\frac{1}{f}\left(-\dot y+\frac{BE\sqrt{\frac{\dot
  y^2}{f}+y^{4-a_1-a_y}}}{\sqrt{1+\frac{B^2E^2}{f}}}\right)
  \ee
with $B=\frac{y^{\frac{1}{2} \left(a_1 (d-1)-\Delta +\Delta _0\right)}}{\rho^{d-2}}$.
Substituting $\dot v$ in (\ref{vdot}) to (\ref{eqmot2}), we obtain the equation of motion for $y(\rho)$  as
  \bea
  \label{eomy}
  &&\ddot y\left(f+E^2B^2y^{2\Delta-2\Delta_0}\right)+\left(f+\dot y^2y^{\Delta}\right)\left(\frac{d-2}{\rho}\dot
  y-\frac{\Delta_0-2\Delta}{2y^{1+2\Delta_0-\Delta}}B^2E^2+\frac{a_1(d-1)}{2y^{1+\Delta}}f\right) \nonumber\\
  &&\quad\quad\quad\quad\quad+\left(B^2E^2y^{\Delta-2\Delta_0}-\dot
  y^2\right)\frac{1}{2}\frac{\partial f}{\partial y}+\frac{\Delta}{2}\left(-y^{-2\Delta_0+\Delta-1}f B^2 E^2+\frac{f\dot
  y^2}{y}\right)=0 \, .
  \eea
Here we define $\Delta=a_1+a_y-4$ to highlight the nonzero $\Delta$ effects as compared to \cite{Fonda} with $\Delta=0$ although in the sphere case the metric has the same spatially isotropic symmetry as in \cite{Fonda}.
The solution of $y(\rho)$ in the $v>0$ and $v<0$ regions are matched at $\rho_c$ with
$v(\rho_c)=0$. Integrating both (\ref{eqmot1}) and (\ref{eqmot2})
across $\rho_c$, the matching conditions are that $\dot v(\rho)$ is continuous across
$\rho_c$ and
  \be
  \label{match}
   \dot y_{+}(\rho_c)=\dot y_{-}(\rho_c)(1-\frac12 g(y_c)) \, ,
  \ee
where $y_c=y(\rho_c)$ and the subscripts $+$ and $-$ denote the
solution in the regions of $v>0$ and $v<0$ separately.
 Once the solution $y(\rho)$ is found,
further integrating $\dot v$ in (\ref{vdot}) with the initial conditions in (\ref{bcyv}) for $v>0$ gives the boundary time
 \be
  \label{time}
  t=\int_{\rho_c}^{R}d\rho\frac{y^{\frac{\Delta_0}{2}}}{h}\left(-\dot y+\frac{BE\sqrt{\dot y^2+h y^{-\Delta}}}{\sqrt{h+B^2E^2}}\right) \, .
  \ee
 In the end, the integral formula
 for the area  (\ref{area}) can be calculated from the contributions of $v<0$ and $v>0$ regions respectively as
\be \label{area_sphere}
\frac{A_\Gamma}{A_{\partial\Sigma}}= \frac{A_{\Gamma v<0}}{A_{\partial\Sigma}}+\frac{A_{\Gamma v>0}}{A_{\partial\Sigma}}=\int_{\rho_c}^{R} \frac{\rho^{d-2}}{y^{\frac{a_1}{2}(d-1)}} \sqrt{{1+\dot y^2 y^{\Delta}}}\,d\rho+\int_{0}^{\rho_c} \frac{\rho^{d-2}}{y^{\frac{a_1}{2}(d-1)}} \sqrt{\frac{h+\dot y^2 y^{\Delta}}{h+B^2E^2}} \,d\rho
\,,
\ee
where we have used the fact that for the $v>0$ region,$
  Q=\frac{h+\dot y^2 y^{\Delta}}{h+B^2E^2}$
by plugging (\ref{vdot}) in (\ref{Q}), and for the $v<0$ region, $Q={1+\dot y^2 y^{\Delta}}$ with $h=1$ and $E=0$.
Translating $\rho_c$ to the boundary time $t$ with the relation (\ref{time}), we are able to find the time-dependent area, and also the corresponding  entanglement entropy.

\subsection{strip}
We now consider the entanglement region $\Sigma$ of a strip extending along say $x_1$-direction from $-R$ to $R$ while other $x_i$'s from $-W$ to $W$ with $W\rightarrow\infty$.
In what follows, we denote  $x_1$ as $x$ for simplifying the notation. In this case, the area of the extremal surface $\Gamma$ is expressed as
\begin{align}\begin{split}\label{25s}
     A_{\Gamma}= A_{\partial \Sigma}\int_{0}^{R}dx\frac{\sqrt{Q}}{y^{\frac{a_1}{2}(\tilde{d}-1)}}
 \end{split} \end{align}
 with
 \begin{equation}\label{qs}
    Q=\Big\vert 1-\frac{\dot{v}^2 f(y,v)}{y^{\Delta_0-\Delta}}-\frac{2 \dot{v} \dot{y}}{y^{\frac{\Delta_0}{2}-\Delta}}\Big\vert \,.
\end{equation}
We define $\dot v=\frac{d}{dx}v(x)$, $\dot
 y=\frac{d}{dx}y(x)$, $A_{\partial \Sigma}=2W^{d-2}$, and $\tilde{d}=\sum_{i=1}^{d-1}\frac{a_i}{a_1}+1$ that can be treated as an effective bulk spacial dimension as also defined in \cite{Giataganas:2013zaa}.

Varying $A_{\Gamma}$ with respect to  the functions $v(x)$ and $y(x)$ leads to  the following equations of motion,
\begin{equation}\label{eqmot1s}
    \sqrt{Q} y^{\Delta_0-\Delta+\frac{a_1}{2}(\tilde{d}-1)}\frac{d}{dx}\left(\frac{y^{-\frac{a_1}{2}(\tilde{d}-1)-\Delta_0+\Delta}}{\sqrt{Q}} \left(\dot{v} f(y,v)+y^{\frac{\Delta_0}{2}} \dot{y}\right)\right)=\frac{1}{2}\dot{v}^2 \frac{\partial f(y,v)}{\partial v} \, ,
\end{equation}
\begin{align}\begin{split}\label{eqmot2s}
    &\sqrt{Q} y^{\Delta_0-\Delta+\frac{a_1}{2}(\tilde{d}-1)}\frac{d}{dx}\left(\frac{ y^{\frac{1}{2} \left(-a_1 (\tilde{d}-1)-\Delta_0+2\Delta\right)}}{\sqrt{Q}} \dot{v}\right)\\
    &=\frac{1}{2}\left(a_1 (\tilde{d}-1) Q y^{\Delta_0-\Delta-1}-\left(\Delta_0-2\Delta\right) y^{\frac{1}{2} \left(\Delta_0-2\right)} \dot{v} \dot{y}+\frac{\dot{v}^2}{y} \left(\left(-\Delta_0+\Delta\right) f(y,v)+y \frac{\partial f(y,v)}{\partial y}\right)\right) \,.
\end{split}\end{align}
The translational symmetry in (\ref{25s}) of $A_{\Gamma}$ in the variable $x$ gives the conserved quantity
 \begin{eqnarray}\label{3.20}
  && y^{\frac{a_1}{2}(\tilde{d}-1)}\sqrt{Q}= J \nonumber\\
 &&\quad\quad \quad = {\rm {constant}}=y_t^{\frac{a_1}{2}(\tilde{d}-1)}\, .
\end{eqnarray}
The value of $J$ can be determined by the boundary condition at $x=0$, which is the tip of the extremal surface  $y(0)=y_t$.
Again, for $\partial_{v}f=0$ in both $v<0$ and $v>0$ regions, there exists another conserved quantity $E$ given by (\ref{eqmot1s}). Together with (\ref{3.20}), we have
 \begin{equation}\label{ess}
y^{-\Delta_0+\Delta} \left(\dot{v} f(y,v)+y^{\frac{\Delta_0}{2}} \dot{y}\right)=E={\rm {constant}}.
\end{equation}
In the vacuum region with $ v<0$, the value of $E$ can be determined at a particular point $x=0$, where the boundary conditions give $E=0$.  Also, $f=1$ in the vacuum region leads to the relation between $\dot{v}$ and $\dot{y}$ at arbitrary $x$ to be
\be \label{60s}\dot{v}=-y^{\frac{\Delta_0}{2}}\dot{y}\,. \ee
Substituting all above relations into (\ref{3.20}), for $v<0$ and  $x>0$ and with no loss of generality, it implies
\be
  \label{slope}
  \dot y=-\frac{\sqrt{y_t^{a_1(\tilde{d}-1)}-y^{a_1(\tilde{d}-1)}}}{y^{\frac{1}{2}(a_1(\tilde{d}-1)+\Delta)}}\, .
   \ee
However, requiring ${dx}/{dy}$ to be finite as $y\rightarrow 0$ gives the constraints
\begin{equation}\label{uv constraint strip}
      a_1(\tilde{d}-1)>0~~~\text{and}~~~ a_1(\tilde{d}-1)\geq-\Delta \, .
\end{equation}
Integrating (\ref{eqmot1s}) and (\ref{eqmot2s}) across the null shell allow us to find the matching conditions, which are the same as those in (\ref{match}) by replacing $\rho$ with $x$. Thus, the matching conditions in this case determine the constant $E$ of (\ref{ess}) for the $v>0$ region in terms of the properties of $y(x)$ at $x_c$, given by
 \begin{equation}\label{ebhs}
{E=\frac{g_c}{2}y_c^{\Delta-\Delta_0/2} \dot{y}_-}\,.
 \end{equation}
Then, via  (\ref{ess}), the relation between $\dot{v}$ and $\dot{y}$ in the black brane region becomes
\begin{align}\begin{split}\label{61s}
\dot{v}=\frac{1}{h}\left(-\dot{y} y^{\frac{\Delta_0}{2}}+E y^{\Delta_0-\Delta}\right)\, .
\end{split}
\end{align}
The equation for $y(x)$
in the black brane region with $v>0$ as in (\ref{slope}) can be found from (\ref{61s}) with the relation (\ref{3.20}) as
\begin{eqnarray}\label{3.36s}
 \dot{y}^{2}&&=H(y)\nonumber\\
 &&=h(y) \Big(\frac{y_{t}^{a_1(\tilde{d}-1)}}{y^{a_1(\tilde{d}-1)}}-1\Big)y^{-\Delta}+E^{2}y^{\Delta_0-2\Delta}\, .
 \end{eqnarray}
 With (\ref{61s}) and the square root of (\ref{3.36s}), we obtain for $x>0$
         \begin{eqnarray}\label{3.37s}
        \frac{dv}{dy}&=& \frac{\dot{v}}{-\sqrt{H(y)}} \nonumber\\
        &=&-\frac{1}{h(y)}\Big(y^{\frac{\Delta_0}{2}}+\frac{Ey^{\Delta_0-\Delta}}{\sqrt{H(y)}}\Big)\, .
        \end{eqnarray}
 In the end, from (\ref{slope}) and (\ref{3.36s}), we can write down the relation between the strip width $R$ and the values of $y_c$ and $y_t$ as
 \be\label{strip R}
 R=\int_{y_c}^{y_t}\frac{y^{\frac{a_1}{2}(\tilde{d}-1)+\frac{\Delta}{2}}}{y_t^{\frac{a_1}{2}(\tilde{d}-1)}\sqrt{1-\frac{y^{a_1(\tilde{d}-1)}}{y_t^{a_1(\tilde{d}-1)}}}}\,dy+\int_{0}^{y_c}\frac{1}{\sqrt{H(y)}}\,dy \,.
 \ee
  From (\ref{3.37s}) and the boundary condition $v(R)=t$, $y_c$ can also be expressed as an implicit function of the boundary time $t$,
 \be\label{time_strip_text}
  t=\int_{0}^{y_c}\frac{1}{h(y)}\Big(y^{\frac{\Delta_0}{2}}+\frac{Ey^{\Delta_0-\Delta}}{\sqrt{H(y)}}\Big)\,dy \, .
 \ee
Moreover,
substituting (\ref{slope}), (\ref{61s}), (\ref{60s}) and (\ref{3.36s}) into (\ref{qs}), we then write the integral formula
 for the area in (\ref{25s}) in terms of the contributions from the $v<0$ and $v>0$ regions respectively as
\be \label{area_strip}
\frac{A_\Gamma}{A_{\partial\Sigma}}= \frac{A_{\Gamma v<0}}{A_{\partial\Sigma}}+\frac{A_{\Gamma v>0}}{A_{\partial\Sigma}}=\int_{y_c}^{y_t}\frac{1}{y^{\frac{a_1}{2}(\tilde{d}-1)-\frac{\Delta}{2}}\sqrt{1-\frac{y^{a_1(\tilde{d}-1)}}{y_t^{a_1(\tilde{d}-1)}}}}\,dy+\int_{0}^{y_c}\frac{y_t^{\frac{a_1}{2}(\tilde{d}-1)}}{y^{a_1(\tilde{d}-1)}\sqrt{H(y)}}\,dy
\, .\ee
Then, through  (\ref{strip R}) and (\ref{time_strip_text}), the area are expressed in terms of the boundary time $t$ and the strip width $R$.
Notice that with the additional constant $J$ in the strip case,  the area in (\ref{25s}) and the corresponding entanglement entropy can be  computed  just by the values of $y(x)$ at $x=0$ and $x=x_c$.

\section{Entanglement Entropy in thermal equilibrium states}\label{sec3}
In this section we consider the entanglement entropy for a final equilibrium state dual in the bulk to a black brane in (\ref{BHef}). We take the large $R$ ($R\gg y_h^{\frac{\Delta}{2}+1}$) and small $R$ ($R\ll y_h^{\frac{\Delta}{2}+1}$) limits respectively, where the former limit corresponds to the case that the tip of the black brane, $y_b$ to be defined later, approaches the black brane horizon, $y_h$ and the latter one is
to assume $y_b \ll y_h$ so that the relevant metric  is that of the pure hyperscaling violating anisotropic Lifshitz spacetime of the form (\ref{metric}) in the case of $h \rightarrow 1$.
The corresponding $A_{\Gamma}$ in both spherical and strip entanglement regions will be computed accordingly.

\subsection{sphere}
 In the case of a spherical $\partial\Sigma$, the extremal surface in the black brane background
(\ref{BHef}) by setting all $a_i$'s to be equal to $a_1$ is denoted as $\Sigma_{BH}$.
The boundary conditions $\dot v(0)=\dot y(0)=0$  give $E=0$ in the
equation of motion (\ref{eomy}). We also denote the solution of the equation
(\ref{eomy}) with $E=0$ and $f=h(y)$, which  satisfies the boundary conditions
(\ref{bcyv}),  by $y_{BH}(\rho)$.  The area of the extremal surface can then be obtained from (\ref{area_sphere}) by setting $\rho_c=R$ and $E=0$ as
\be\label{area_eq_sphere}
A_{\Gamma eq}=\frac{A_{\partial
   \Sigma}}{R^{d-2}}\int_0^{R}\frac{\rho^{d-2}}{y_{BH} (\rho)^{\frac{a_1}{2}(d-1)}}\sqrt{1+\frac{\dot{y}_{BH} (\rho)^2y_{BH} (\rho)^\Delta}{h(y_{BH} (\rho))}}\,d\rho \, .
\ee
We first consider the extremal surface in the large $R$ limit ($R\gg y_h^{\frac{\Delta}{2}+1}$).
In this case, it is anticipated  that the tip of the $\Sigma_{BH}$, denoted
as $y_{BH}(0)\equiv y_b$, is very close to the horizon $y_h$, namely
  \be\label{ybyh_text}
  y_b\simeq y_h(1-\epsilon) \, ,
  \ee
where $\epsilon\ll 1$.  The solution of $y_{BH}$ near the horizon can then be approximated by
\be \label{yBH}
y_{BH} (\rho)\simeq y_h- y_1(\rho)\epsilon+O(\epsilon^2)
\ee
with the first order perturbation $y_1(\rho)$ given by
\be\label{y1_text}
y_1(\rho)=\Gamma^2(\frac{d-1}{2})\left(\frac{\rho\gamma}{2}\right)^{3-d}I^2_{\frac{d-3}{2}}(\gamma \rho) \, ,
\ee
{where $I_\mu(x)$ is the modified Bessel function of the first kind.
Let us
denote the inverse function of $y_{BH}(\rho)$ as $\rho_{BH}(y)$.
 We also have the expansion of $\rho_{BH}(y)$ near the boundary $y=0$ as
\begin{equation} \label{y_BH}
\rho_{BH}(y)=R-\rho_0(y)+O(1/R)\, ,
\end{equation}
{where}
\be\label{rho0_text}
\rho_0(y)=\int_0^y\frac{\psi^{\frac{1}{2}\left(\Delta+a_1(d-1)\right)}}{\sqrt{{y_b}^{a_1(d-1)}-\psi^{a_1(d-1)}}\sqrt{h(\psi)}} \,d\psi \, .
\ee

To find the relation between $\epsilon$  and $R$, the existence of a matching region for the above two solutions is crucial \cite{RG}.
We extend the solution (\ref{y1_text}) (near the horizon) to the region of relatively large $\rho$ where
 $\rho\ll R-O(R^0)$ is still satisfied, and then extend the solution (\ref{rho0_text}) (near the boundary)
to the region of $\epsilon \ll 1-y/y_h \ll 1$ where $ R-\rho \ll O(R^1)$ is still valid for being consistent with (\ref{y_BH}).  We find that
 with such extensions, both solutions (\ref{yBH}) and (\ref{y_BH}) have the similar structure $(R-\rho) \propto -\ln(y_h-y) $ in the matching region where the relation of $\epsilon$ and $R$ can be read off by comparing their  leading order behaviour given by}
  \be \label{epsilon_R}
   \epsilon \simeq e^{-2 \gamma R}
  \ee
with $-2\gamma^2 \equiv
\frac{h'(y_h)}{y_h^{\Delta+1}}\frac{a_1}{2}(d-1)$.
Apparently, the large $R$ limit drives  $\epsilon$ to a small value. The nonzero $\Delta$ contributes to the value of $\gamma$ in (\ref{epsilon_R}), which is different from the one in AdS spacetime \cite{Liu-s}.

The area of the extremal surface in the large $R$ limit can be computed from the solutions (\ref{yBH}) and (\ref{y_BH})  by splitting the integral (\ref{area_eq_sphere})
into the areas near the horizon (IR) and near the boundary (UV).  Then the UV divergence part of $A_{\Gamma eq}$ is
  \be
 A_{\Gamma  div}\simeq \frac{2A_{\partial
   \Sigma}}{a_1(d-1)-(2+\Delta)}\left(\frac{1}{y_{UV}^{\frac{1}{2}(a_1(d-1)-2-\Delta)}}\right)\, ,
\ee
where $y_{UV}$ is a UV-cutoff in the $y$ integration.  The finite part  is obtained as
\be
\Delta A_{\Gamma }=A_{\Gamma}-A_{\Gamma div}=\frac{V_{\partial
   \Sigma}}{y_h^{\frac{a_1}{2}(d-1)}}+O(R^{d-2})
\ee
where $V_{\partial
\Sigma}=\frac{\pi^{\frac{d-1}{2}}}{\Gamma(\frac{d+1}{2})}R^{d-1}$ is the volume of a $(d-1)-$ball with radius $R$.
Similar results are also obtained in \cite{Fonda} for $\Delta=0$.

In the small $R$ limit, namely ($R\ll y_h^{\frac{\Delta}{2}+1}$), the area of the extremal surface is determined by (\ref{area_eq_sphere}) in the limit of $h \rightarrow 1$.
In particular, for $a_y=2$ we can find an exact solution of (\ref{eomy}),
  \be\label{exact_text}
  \rho^{(0)}(y)=\sqrt{R^2-\frac{4}{a_1^2}y^{(0) a_1}} \, ,
  \ee
where the relation of the tip of the extremal surface $y_b^{(0)}$ and $R$, an essential information to analytically find the finite part of the area, can be read off as
\begin{equation} \label{yb_R}
 y_b^{(0)}=y^{(0)}(0)=
 \left(\frac{a_1^2 R^2}{4}\right)^{1/a_1} \, .
\end{equation}
Substituting (\ref{exact_text}) into (\ref{area_eq_sphere}) ($h \rightarrow 1$), and again dividing the area into divergent and finite parts, we have
 \be \label{A_0} A^{(0)}_{\Gamma eq}=A^{(0)}_{\Gamma div}+ \Delta A^{(0)}_{\Gamma eq}  \ee
 where
 \be
A^{(0)}_{\Gamma div}  = \frac{2A_{\partial \Sigma}}{a_1(d-2)y_{UV}^{\frac{a_1}{2}(d-2)}}-\frac{4(d-3)A_{\partial \Sigma} R^{-2}}{a_1^3(d-4)y_{UV}^{\frac{a_1}{2}(d-4)}}+...+
\begin{cases}
        (-1)^{\frac{d}{2}}A_{\partial \Sigma} \frac{2^{d-1}(d-3)!!}{a_1^{d-1}(d-2)!!}R^{-d+2}\ln\left({y_{UV}}\right)& \text{$d$ is even}\\
        (-1)^{\frac{d+1}{2}}A_{\partial \Sigma}\frac{2^{d-2}R^{-d+3}}{a_1^{d-2}y_{UV}^{a_1/2}}& \text{$d$ is odd}.        \end{cases}
\ee
 \be
 \Delta A^{(0)}_{\Gamma eq}=\frac{A_{\partial \Sigma}}{R^{d-2}}
 \begin{cases}
        (-1)^{\frac{d}{2}+1} \frac{2^{d-1}(d-3)!!}{a_1^{d-1}(d-2)!!}\ln\left(R\right) +O(R^0/y_{UV}^0)& \text{$d$ is even}\\
        \frac{2^{d-2}\Gamma(1-\frac{d}{2})\Gamma(\frac{d-1}{2})}{\sqrt{\pi} a_1^{d-1}}& \text{$d$ is odd}.        \end{cases}
\ee
$A_{\partial \Sigma}=\frac{2\pi^{\frac{d-1}{2}}}{\Gamma(\frac{d-1}{2})}R^{d-2}$ is the area of a $d-2$-sphere with radius
R. Here we restrict ourselves to  $d\ge 3$ so that the dimension of $\partial\Sigma$ is larger than or equal to 2. The above results can reproduce the ones in AdS space \cite{Ryu-d} by choosing appropriate values of the scaling parameters. Although $a_y=2$ is chosen, our results with $\Delta \ne 0$ generalize the result of \cite{Fonda}. Through  the Ryu-Takayanagi
formula (\ref{S_A}), the corresponding entanglement entropy can also be obtained where its finite part is important to make a comparison with
the field theory results.

\subsection{strip}
 To find the area of the extremal surface for the strip case in the thermal equilibrium, we substitute the relation between $R$ and $y_b$  in (\ref{strip R}) to (\ref{area_strip}) by setting  $y_c=y_t=y_b$ and $E=0$, we have
 \be \label{area_eq_strip}
{A_{\Gamma eq}}={A_{\partial\Sigma}} \int_{0}^{y_b}\frac{y_b^{\frac{a_1}{2}(\tilde{d}-1)}}{y^{a_1(\tilde{d}-1)}\sqrt{H(y)}}\,dy \, ,
\ee
where $H(y)$ is defined in (\ref{3.36s}).
In the large $R$ limit, the tip of the extremal surface $y_b$ is assumed to be close to $y_h$ in terms of the expansion of (\ref{yBH}) for small $\epsilon$ where again $\epsilon$ can be related to $R$  by (\ref{epsilon_R}).
The straightforward calculations show that the UV divergent part of the area is
\be \label{A_div}
A_{\Gamma div}=\frac{2A_{\partial
   \Sigma}}{a_1(\tilde{d}-1)-(2+\Delta)}\left(\frac{1}{y_{UV}^{\frac{1}{2}(a_1(\tilde{d}-1)-2-\Delta)}}\right) \, ,
   \ee
and the finite part is
\be\label{large R eq A}
A_{\Gamma eq}-A_{\Gamma div}\simeq \frac{A_{\partial\Sigma} R}{y_h^{\frac{a_1}{2}(\tilde{d}-1)}} \, .
\ee
Note that $A_{\Gamma div}$ vanishes when $a_1(\tilde{d}-1)<2+\Delta $ as  $y_{UV} \rightarrow 0.$
The contribution from the scaling parameter $a_1$  to the area of the extremal surface plays the same role as in the sphere case, leading to the same entanglement entropy for both the sphere and strip cases in our setting.
Nevertheless, it will be seen that the subsequent time evolution after a quench for two cases
are very different, in particular during the late-time thermalization processes.

In the small $R$ limit, from (\ref{strip R}) with $y_t\rightarrow y_b^{(0)}$ and $y_c \rightarrow 0$, we obtain the relation between $R$ and $y_b^{(0)}$ as
\be\label{strip R small_text}
R\simeq \tilde{\mathscr{R}}_0 \, {a_1(\tilde{d}-1)}y_b^{\frac{\Delta}{2}+1}
\ee
with $\tilde{\mathscr{R}}_0=\sqrt{\pi} \Gamma(\frac{1}{2}+\frac{\Delta+2}{2a_1(\tilde{d}-1)})\Gamma^{-1}(1+\frac{\Delta+2}{2a_1(\tilde{d}-1)})$ a dimensionless constant. In particular,
\begin{equation}\label{delta_c}
\Delta >-2
\end{equation}
is required for a sensible result.
The area of (\ref{area_eq_strip})
for $a_1(\tilde{d}-1)\neq 2+\Delta $ becomes
\be
A_{\Gamma eq}-A_{\Gamma div}\simeq A_{\partial\Sigma} \frac{y_b^{(0) \frac{\Delta}{2}+1-\frac{a_1}{2}(\tilde{d}-1)}\sqrt{\pi}}{a_1(\tilde{d}-1)}\frac{ \Gamma\left(-\frac{1}{2}+\frac{\Delta+2}{2a_1(\tilde{d}-1)}\right)}{\Gamma\left(\frac{\Delta+2}{2a_1(\tilde{d}-1)}\right)} \,
\ee
with $A_{\Gamma div}$ in (\ref{A_div}).
However, for $a_1(\tilde{d}-1)= 2+\Delta $
\be\label{strip zero small ln}
A_{\Gamma eq}\simeq A_{\partial\Sigma}\ln \Big(\frac{y_b^{(0)}}{y_{UV}}\Big) \, .
\ee
The critical value determined by $a_1(\tilde{d}-1)= 2+\Delta $ with the logarithmic divergence rather than the power-law ones generalizes the result in \cite{Fonda} where an isotropic background is considered.
In particular, when $a_y=2$ and all $a_i$'s equal to $a_1$, the area of the extremal surface still has quite different behavior from that of the sphere case in the small $R$ limit.

Starting from the next section, we will focus on the nonequilibrium aspect of
thermalization processes, which is encoded in the time-dependent entanglement entropy. It is known from previous studies  that the black brane horizon radius $y_h$ sets a time scale for the nonequilibrium  system to reach the ``local equilibrium" as to cease the production of thermodynamical entropy.
In the large $R$ limit with $R\gg y_h^{\frac{\Delta}{2}+1}$, thermodynamical entropy of a system generally evolves through "pre-local" equilibrium growth in the early times $t\ll y_h^{\frac{\Delta}{2}+1}$, the linear growth in the intermediate times when $R \gg t\gg y_h^{\frac{\Delta}{2}+1}$, and the final saturation stage when $t\rightarrow t_s \propto R$.  On the contrary, in the small $R$ limit with $R\ll y_h^{\frac{\Delta}{2}+1}$, it is anticipated that after the saturation, the tip of the extremal surface $y_t$ in the end is still far away from the horizon $y_h$, namely $y_t \ll y_h$. Thus, after the "pre-local" equilibrium growth, the system will directly reach the saturation stage near the saturation time scale, $t_s$ determined by the size of the system $R$,  which will be studied later. For a quantitative comparison of  the  entanglement entropy between the small and large $R$ limits, we just consider the time-dependent entanglement entropy during the early time growth and the final saturation stage in the following sections.

\section{The early time entanglement entropy growth}\label{sec4}
In this section, we study the entanglement growth at the early times when the infalling shell meets the extremal surface at $y_c$ with the condition that $y_c^{\frac{\Delta}{2}+1} \ll R$ for both large and small $R$ limits.
This means that the infalling shell is very close to the
boundary $y=0$ so that the extremal surface $\Gamma$ is mostly in
the $v<0$ region with the pure hyperscaling violating anisotropic Lifshitz metric.  During such early times, the infalling shell does not have much enough
 time to probe the whole geometry so we expect that the same growth rate will be found for both the large and small $R$ limits in either the sphere or the strip case.

\subsection{sphere}
 Here we start by considering the zeroth order solution of $\Gamma$ that satisfies the equation of motion for $y$ in (\ref{eomy}) with $f=1$ and $E=0$. Let $y^{(0)}(\rho)$ be the solution of this equation with the
boundary condition $\dot y^{(0)} (0)=0$ and $y^{(0)} (R)=0$, and
$\rho^{(0)}(y)$  be the inverse function of $y^{(0)}(\rho)$.
The superscript  $(0)$ means the extremal surface dual to the vacuum state of the system.
The area of the extremal surface is given in (\ref{A_0}).
In the early times, the infalling shell intersects $\Gamma$ at a place with the value of
 $ y_c^{\frac{\Delta}{2}+1}\ll R$ in both small and large $R$ cases.
For such a small $y_c$, the relevant zeroth order solution near the boundary for small $y$ is obtained from (\ref{y_BH}) and (\ref{rho0_text}) as
  \be
  \label{yb}
  \rho^{(0)}(y)=R+Cy^{\Delta+2}+O(y^{2\Delta+3})
  \ee
with
$C=-\frac{2(d-2)}{R}\left((\Delta +2) \left(a_1 (d-1)-\Delta -2\right)\right)^{-1}$.
Also,  when
\begin{equation}\label{h_v_c_sphere}
\Delta +2 < {\Delta_h} -1 \, ,
\end{equation}
the blackening factor can be safely  approximated by
$h(y) \rightarrow 1$ for small $y$. The existence of  the extremal surface holomorphic to $\Sigma$
requires that $C>0$ and the finiteness of ${d\rho}/{dy}$ as $y\rightarrow 0$ gives further constraints on the scaling parameters
\begin{equation}\label{a1_c}
\Delta>-2 \, ,~~~ a_1>\frac{\Delta+2}{d-1}>0 \, .
\end{equation}
Then, we rewrite the area integral  (\ref{area})
in terms of the variable $\rho(y)$, which is the inverse function of $y(\rho)$,
as
\be
 A_{\Gamma}(t)=\frac{A_{\partial
   \Sigma}}{R^{d-2}}\int_0^{y_t}dy\frac{\rho^{d-2}}{y^{\frac{a_1}{2}(d-1)}}\sqrt{\rho'^2-y^{\Delta-\Delta_0}fv'^2-2y^{\Delta -\frac{\Delta _0}{2}}v'}
   \ee
where the prime means the derivative with respect to $y$ and $y_t$ is the tip of the extremal surface, namely, $y_t=y(0)$.
For the early times with small $t$, let us consider the perturbations around the zeroth order solution where
the time-dependent  $A_{\Gamma}(t)$ just slightly departs from $A^{(0)}$ as
  \be
  \label{variate}
  \delta A_{\Gamma}(t)\equiv A_{\Gamma}(t)-A^{(0)}\simeq \frac{\partial A_{\Gamma}(t)}{\partial
  y_t}\delta y_t+\frac{\partial A_{\Gamma}(t)}{\partial
  v}\delta v+\frac{\partial A_{\Gamma}(t)}{\partial
  \rho}\delta \rho+\frac{\partial A_{\Gamma}(t)}{\partial
  f}\delta f \, .
  \ee
The partial derivatives are evaluated at $ y=y^{(0)}$ or $\rho=\rho^{(0)}$ and $f=1$ where
$y_t=y^{(0)}(0)$ and $v=v(y^{(0)})=v^{(0)}$. The first term vanishes due to the vanishing
of the area at the tip $y=y^{(0)}$. The second and the third terms
vanish due to the equations of motion for $\rho$ and $v$. For the last term, we have
 \be \label{df}
 \delta f=h(y)~~\mbox{for $y<y_c$},~~~\delta f=0~~\mbox{for $y>y_c$} \, .
 \ee
  $y_c$ is the position where the infalling sheet intersects the extremal surface, namely $ v^{(0)}(y_c)=0$.
Since $y_c\ll y_h$, the near boundary behavior for
$h(y)$ in (\ref{BHb}) is applied.
Solving  (\ref{vdot}) in the case of
$E=0$ and $f=1$, and with the boundary conditions (\ref{bcyv}),
$
  v^{(0)}(y)\simeq t-\frac{y^{\frac{\Delta_0}{2}+1}}{\frac{\Delta_0}{2}+1}
$
near $y=0$.  Putting all together, we can find $\delta A_{\Gamma}$ in (\ref{variate})
as a function of $y_c$.
As a result of $v(y_c)=0$, the $y_c$ dependence can be translated into that of the boundary time $t$ by
$t\simeq\frac{y_c^{\frac{\Delta_0}{2}+1}}{\frac{\Delta_0}{2}+1}$. Note that
\begin{equation} \label{delta'_c}
\Delta_0>-2
 \end{equation}
 is required so that the boundary time $t$ increases in $y_c$.  Thus, in terms of the boundary time, the early time entanglement entropy growth for both small and large $R$ limits is given by
  \be\label{early growth}
\delta S(t)\simeq \frac{\delta A_{\Gamma}(t)}{4}\simeq \frac{A_{\partial
\Sigma}M}{4(\Delta+2{\Delta_h}+2-a_1(d-1))}\left(\frac{\Delta_0}{2}+1\right)^{\frac{\Delta+2{\Delta_h}+2-a_1(d-1)}{\Delta_0+2}}t^{\frac{\Delta+2{\Delta_h}+2-a_1(d-1)}{\Delta_0+2}}
\,  \ee
where
\begin{equation} \label{S_c_sphere}
\Delta+2{\Delta_h}+2-a_1(d-1) >0
\end{equation}
to ensure that the entanglement entropy increases in time.
The extra $\Delta$ dependence of the growth rate of the entanglement entropy generalizes the results  in \cite{Fonda} with $\Delta=0$.
In particular, for the positive (negative) value of $\Delta$, the power of $t$ increases (decreases) with $\Delta$  ($ |\Delta|$) that speeds up (slows down) the growth rates as compared with the case of $\Delta=0$.

\subsection{strip}
In the strip case, the area of the extremal surface can be obtained from (\ref{area_strip}) by setting $y_c=y_t$, giving
\be\label{strip_area_parametrized_by_y}
 A_{\Gamma}(t)=A_{\partial
   \Sigma}\int_0^{y_t}dy\frac{1}{y^{\frac{a_1}{2}(\tilde{d}-1)}}\sqrt{x'^2-y^{\Delta-\Delta_0}fv'^2-2y^{\Delta -\frac{\Delta _0}{2}}v'}
   \ee
where again the prime means the derivative with respect to $y$.
As in the sphere case, the non-vanishing term in the variation of $ A_{\Gamma}(t)$ evaluated at the zeroth order solutions $y^{(0)}$ and $v^{(0)}=v(y^{(0)})$ in  (\ref{slope}) and (\ref{efcoor}), is
\be
  \label{variate_s_test}
  \delta A_{\Gamma}(t)\simeq \Eval{\frac{\partial A_{\Gamma}}{\partial f}}{(0)}{}\delta f \, .
  \ee
With $\delta f$ in (\ref{df}), straightforward calculations
give the same entanglement entropy growth at the early times as in (\ref{early growth}) by replacing $d$ with $\tilde{d}$.
As the comparison to \cite{Fonda} with $\Delta=0$, the nontrivial dependence of $\Delta$ in the early time growth rate
can in principle be tested experimentally.

 \section{The late time saturation} \label{sec5}
Although the entanglement entropy in the early times exhibits the same growth rate in both the sphere and strip cases, the late time thermalization process will find different time dependent behaviors for two cases. Due to the fact that the entanglement entropy growth can be realized as an ``entanglement tsunami" led by a sharp front moving inward from the boundary $\Sigma$ \cite{Liu-s}, the saturation behavior will be the same but the saturation time scales might be very different.
 Also, for a given geometry, we will work on the large and small $R$ limits separatively.
\subsection{sphere}
Let us start with the sphere case in the large $R$ limit.
Near the saturation, the extremal surface $\Gamma$  is
mostly in the $v>0$ black hole region, and  will become very
close to the one in the purely black hole background
(\ref{BHef}).
Thus, the infalling shell, which is very near the
 tip of $\Gamma$ at $\rho=0$, can be parametrized as
 \be
   \label{delta}
   \rho_c=y_c^{\frac{\Delta}{2}+1}\delta
 \ee
with $\delta\ll 1$. Also, near $\rho=0$ for $v<0$, the extremal
surface $y(\rho)$ satisfies (\ref{eomy}) with $E=0$ and $ f=1$ where the leading order solution $y(\rho)$ can be approximated by
\be
\label{ytip}
y(\rho)\simeq y_t-\frac{a_1}{4}y_t^{-\Delta-1}\rho^2 \, .
\ee
From (\ref{delta}) and the definition $y_c=y(\rho_c)$, the relation between the tip of the extremal surface $y_t$ and the infalling sheet $y_c$ is obtained as
  \be
  \label{yt}
  y_t\simeq y_c(1+\frac{a_1}{4}\delta^2) \,.
  \ee
Also, from the matching condition (\ref{match}) at $\rho^* \approx \rho_c$ and the
approximation solution (\ref{ytip}), the variables $y(\rho)$ and $v(\rho)$ at the matching point are
given respectively by
  \be
  \label{bctip}
   \dot{y}_-\simeq-\frac{a_1}{2}\frac{\rho_c}{y_c^{\Delta+1}},~~\dot{y}_+=\dot{y}_-(1-\frac12
  g(y_c)),~~\dot{v}_+=\dot{v}_-=-y_c^{\frac{\Delta_0}{2}}\dot{y}_- \, .
  \ee
Plugging them into (\ref{es}), the constant of motion $E$ for $v>0$ in the
black hole region is found to be
  \be
  \label{Eap}
 E\simeq -\frac{a_1 g(y_c)}{4}\delta^{d-1}y_c^{\frac{1}{2} \left((d-1) \left(-a_1+\Delta +2\right)-\Delta _0-2\right)} \, .
  \ee
Next, we also expand $y(\rho)$ around
$y_{BH}(\rho)$ in the $v>0$ region, which in the small $\rho$ approximation is given by
\be
  \label{yrho}
 y(\rho)= y_{BH}(\rho)+ \frac{a_1 g(y_b)}{2(d-3)}y_b^{(\frac{\Delta}{2}+1)(d-1)-1}\delta^{d-1}\rho ^{-d+3}+O\left(\delta^{2(d-1)}\right) \, .
\ee
In the sphere case, $d >3$ is considered.
Given $y(\rho\rightarrow 0)=y_t$ and $y_{BH}(\rho \rightarrow 0)=y_b$ in (\ref{yrho}), the relation between $y_t$ and $y_b$ can be obtained
 from (\ref{yt}) as
 \be
 \label{ybyc}
y_c=y_b(1+c_2 \delta^2+O(\delta^4))~~~
 \ee
with $c_2=\frac{a_1}{4}\left(\frac{(d-2)g(y_b)}{(d-3)}-1\right)$ where the boundary time $t$ depends on the infalling sheet $y_c$ through (\ref{time}).

 We then calculate the time dependent entanglement entropy near the saturation after a quench. The area of the extremal surface can be divided into the $v<0$ and
$v>0$ parts, $A_{\Gamma}=A_{\Gamma v<0}+A_{\Gamma v>0}$. Near the saturation, the solution around $\rho=0$ in (\ref{ytip}) for $v<0$ and (\ref{yrho}) for $v>0$  will be relevant.  We denote $\Delta A_{\Gamma}=A_{\Gamma v<0}+A_{\Gamma v>0}-A_{\Gamma eq}$, by subtracting the extremal surface in  thermal equilibrium due to the pure black hole in (\ref{area_eq_sphere}) where $\Delta A_{\Gamma}$  becomes
  \be
  \label{deltaA}
 \Delta A_{\Gamma}\simeq -\frac{A_{\partial \Sigma }}{R^{d-2}}\frac{a_1^2 g\left(y_b\right)}{8 (d-3)}\left(\frac{g\left(y_b\right)}{4 h\left(y_b\right)}+\frac{d-3}{d+1}\right) y_b^{\frac{1}{2} (d-1) \left(-a_1+\Delta +2\right)}\delta ^{d+1} \, .
  \ee
From (\ref{time}), the boundary time can be written as  $t=t_1+t_2$ given by
\be\label{t1t2}
   t_1=\int_0^{y_c}dy\frac{y^{\frac{\Delta_0}{2}}}{h(y)},~~~t_2=E\int_{\rho_c}^Rd\rho\frac{y^{\frac{\Delta_0}{2}}}{h(y(\rho))}\frac{y^{\frac{a_1}{2}(d-1)+\frac{\Delta_0}{2}-\frac{\Delta}{2}}}{\rho^{d-2}}\frac{\sqrt{\dot y^2+h(y(\rho))y^{-\Delta}}}{\sqrt{h(y(\rho))+B^2E^2}} \, .
 \ee
We can evaluate the saturation time, $t_s$ in the large $R$ limit by (\ref{vdot}) with $E=0$, $f=h$ and the conditions
$v_{BH}(R)=t$ and  $v_{BH}(y_b)=0$ at $t=t_s$, given by
  \be
  \label{ts}
t_s=\int_0^{y_b}dy \frac{y^{\frac{\Delta_0}{2}}}{h(y)}=\frac{R}{c_E}-\frac{d-2}{4\pi T}\ln R+O(R^0)~~~\text{with $c_E=\sqrt{\frac{4\pi T y_h^{\Delta+1-\Delta_0/2}}{a_1(d-1)}}$} \, .
  \ee
Here $T$ is the black brane temperature defined above, and the  above integral is dominated in the near horizon region.
In terms of $T$,   $c_E \propto T^{\frac{\Delta_0-\Delta}{\Delta_0+2}}$. The saturation time $t_s \propto R$ in the large $R$ limit.
 For a positive (negative) $\Delta$,
$c_E$ becomes smaller (larger) than the case of $\Delta=0$, leading to relatively larger (smaller) the saturation time scale $t_s$ for  fixed $R$ and $T$.

The integral of $t_1$ can be further separated into two parts as
$
  t_1=t_s+\int_{y_b}^{y_c}dy\frac{y^{\frac{\Delta_0}{2}}}{h(y)} \, .
$
Then, near the saturation as $\delta\rightarrow 0$ and with the
approximate solutions (\ref{yrho}), (\ref{ybyc}) and (\ref{Eap}),
we find, to leading order in $\delta$,
  \be \label{t-ts}
  t\simeq t_s-\frac{a_1}{4}y_b^{\frac{\Delta_0}{2}+1}\delta^2 \,.
  \ee
Together with (\ref{deltaA}) and through (\ref{S_A}) give
  \be \label{deltaA_s_text}
\Delta S (t)= \frac{\Delta A_{\Gamma}}{4} \propto (t-t_s)^{\frac{d+1}{2}}~ \, .
  \ee
Note that  $d\geq 4$ is required. Based upon the constraints in (\ref{bk_c}) from the property of the black brane,  (\ref{nec1}),(\ref{nec2}), and (\ref{nec3}) due to null energy conditions, and (\ref{h_v_c_sphere}) and  (\ref{a1_c}) from the solution of the extremal surface as well as
(\ref{delta'_c}) and (\ref{S_c_sphere}) for the sensible entanglement entropy, $a_1$ is always positive so that
the boundary time $t$ approaches to $t_s$ from below. Thus,
the saturation can  be reached continuously. It will be compared with the strip case where the continuous saturation will occur only for some parameter regions to be discussed later.
The saturation behaviour of the entanglement entropy does not depend on the nonzero $\Delta$
 in the sense that the power law saturation depends  only on the dimension $d$.
 Nevertheless, the saturation time scale $t_s$ depends on the nonzero value of $\Delta$ as expected.

In the small $R$ limit, the analysis of saturation behaviors will be different from the large $R$ case to be explained as follows.
In the large $R$ limit,  the tip of the
extremal surface in the pure black hole geometry, $y_b$ is
exponentially close to the horizon $y_h$ where  $h(y_b)\propto e^{-2\gamma R}$. However,  in the small $R$ limit, $y_b \ll y_h$ giving $h(y_b) \simeq 1$.
Moreover, we should have the same relation between $y_b$ and $y_c$ as in the large $R$ limit (\ref{ybyc}) but with the different $d_1$ to account for the fact that $h(y_b) \simeq 1$.
Also, in the small $\delta$ expansion of $\Delta A_{\Gamma}$, the power of $\delta$ in the leading order is the same as in (\ref{deltaA}). In the end, the same saturation behavior as in ({\ref{deltaA_s_text}})  is found in the small $R$. The saturation time $t_s$ is different from what is obtained in the large $R$ limit. Having the analytical expression of $t_s$ as a function of $R$ resides in the exact solution of $y(\rho)$ in pure hyperscaling violating Lifshitz spacetime with anisotropic scalings in spatial coordinates.
An exact solution can be found for  $a_y=2$ as we have discussed in (\ref{exact_text}). With the relation of $y_b$ and $R$ in (\ref{yb_R}), (\ref{ts}) instead gives,
  \be
 t_s\simeq \frac{2}{a_0}\left(\frac{a_1^2 R^2}{4}\right)^{\frac{a_0}{2 a_1}}\,
  \ee
for small $R$. This nontrivial power-law dependence of $R$ on the saturation time scale $t_s$ brings in an interesting
probe of the systems toward thermalization with different sizes $R$. Moreover for small $R$, since $a_1$ is always positive,
the saturation time $t_s$ becomes larger (smaller) as  $a_1$ increases (decreases).

\subsection{strip}\label{EE in strip late time_text}
 In the strip case, the relation between $x_c$ and $y_c$ near the saturation is the same as in (\ref{delta}) by replacing $\rho_c$ by $x_c$.
 From (\ref{slope}), we are able to find the solution near the IR region for $v<0$
  \be\label{strip ir vac_text}
  y(x)=y_t-\frac{a_1(\tilde{d}-1)}{4y_t^{\Delta +1}}x^2+O(x^4) \, .
  \ee
  Due to the relation between $x_c$ and $y_c$, and the definition of $y_c=y(x_c)$,  from (\ref{strip ir vac_text}) we obtain
  \be\label{ycyt_text}
  y_c=y_t\left(1-\frac{a_1}{4}(\tilde{d}-1)\delta^2+O(\delta^4)\right) \, .
  \ee
  Also, from (\ref{ycyt_text}) and  (\ref{slope}), the conserved quantity $E$  (\ref{ebhs}) in the region $v>0$ can be approximated in terms of the small $\delta$ by
  \be
  E\simeq -\frac{a_1}{4}(\tilde{d}-1)g(y_t)y_t^{\frac{\Delta}{2}-\frac{\Delta_0}{2}}\delta \, .
  \ee
  Recall that in the large (small) $R$ limit, $h(y_b)\simeq {\Delta_h} e^{-2\gamma R}$ ($h(y_b)\simeq 1$).
  We then assume the relation between $y_b$ and $y_t$ to be
  \be y_t=y_b(1+\xi)\ee
 with a small parameter $\xi$.
 To find the relation between $\xi$ and $\delta$,  we rewrite (\ref{strip R}) as
  \be\label{Rrerite_text} R=R_1-R_2+R_3+\boldsymbol{\mathscr{R}}(y_t)   \, ,\ee
  where
  \be\label{r1_r2_r3_text} R_1\equiv\int_{y_c}^{y_t}\frac{y^{\frac{\Delta}{2}}dy}{\sqrt{\frac{y_t^{a_1(\tilde{d}-1)}}{y^{a_1(\tilde{d}-1)}}-1}},~~~R_2\equiv\int_{y_c}^{y_t}\frac{dy}{\sqrt{H(y)}},~~~R_3\equiv\int_{0}^{y_t}\left(\frac{1}{\sqrt{H(y)}}-\frac{y^{\frac{\Delta}{2}}}{\sqrt{h(y)\left(\frac{y_t^{a_1(\tilde{d}-1)}}{y^{a_1(\tilde{d}-1)}}-1\right)}}\right)dy\ee
  and the function $\boldsymbol{\mathscr{R}}$ is defined by
  \be\label{r_scr_text}
 \boldsymbol{\mathscr{R}}(y_t)\equiv \int_{0}^{y_t}\frac{y^{\frac{a_1}{2}(\tilde{d}-1)+\frac{\Delta}{2}}}{y_t^{\frac{a_1}{2}(\tilde{d}-1)}\sqrt{h(y)}\sqrt{1-\frac{y^{a_1(\tilde{d}-1)}}{y_t^{a_1(\tilde{d}-1)}}}}\,dy \, .
  \ee
Also, notice that $R$ can be expressed at the thermal equilibrium by
$R=\boldsymbol{\mathscr{R}}(y_b)$.
The integration (\ref{r1_r2_r3_text}) can be expanded in the small $\delta$, and to the leading order they are $ R_1\simeq R_2 \simeq y_t^{1+\frac{\Delta}{2}}\delta$, and $R_3\simeq \frac{2y_t^{1+\frac{\Delta_0}{2}}E}{a_1(\tilde{d}-1)h(y_t)} \, .$
Moreover, we expand $\boldsymbol{\mathscr{R}}(y_t)$ at $y_t=y_b$ for the small $\xi$ as
$
\boldsymbol{\mathscr{R}}(y_t)=\boldsymbol{\mathscr{R}}(y_b)+\boldsymbol{\mathscr{R}}'(y_b) y_b \, \xi +O(\xi^2)\, .
$
With  their expansions for small $\xi$ and $\delta$, from (\ref{Rrerite_text}) and (\ref{r_scr_text})  the relation between $\xi$ and $\delta$ is found to be
 {$
 \xi\simeq \frac{g(y_b)y_b^{\frac{\Delta}{2}}}{2\boldsymbol{\mathscr{R}}'(y_b)h(y_b)}\delta \, .
 $}

We then divide the boundary time $t$ (\ref{time_strip_text}) into four parts
\be
{t=t_s+t_1+t_2-t_3} \, ,
\ee
where
\be\label{t1_t2_t3_text}
t_1\equiv\int_{y_b}^{y_c}\frac{y^{\frac{\Delta_0}{2}}}{h(y)}\,dy, ~~~ t_2\equiv E\int_{0}^{y_t}\frac{y^{\Delta_0-\Delta}}{h(y)\sqrt{H(y)}}\,dy,~~~t_3\equiv E\int_{y_c}^{y_t}\frac{y^{\Delta_0-\Delta}}{h(y)\sqrt{H(y)}}\,dy
\ee
and $t_s$ is the saturation time, obtained from (\ref{time_strip_text}) by setting $y_c=y_b$ and $E=0$.
In the large $R$ limit, since the relation between $y_b$ and $R$ is identical to that in the sphere case in (\ref{ybyh_text}) and (\ref{epsilon_R}), the leading behaviour of $t_s$ is also the same as (\ref{ts}).
For a positive (negative) $\Delta$, the saturation time scale $t_s$ for  fixed $R$ and $T$
 becomes larger (smaller) than the case of $\Delta=0$.
In the small $R $ limit, the relation between the tip of the extremal surface $y_b$ and the size of the boundary $R$ in (\ref{strip R small_text}) allow us to write down the approximate saturation time $t_s$
\be\label{ts strip smallr_test}
t_s\simeq \frac{2}{\Delta_0+2}\left(\frac{a_1(\tilde{d}-1)}{\tilde{\mathscr{R}}_0}\right)^{\frac{\Delta_0+2}{\Delta+2}}R^{\frac{\Delta_0+2}{\Delta+2}} \,
\ee
with ${\mathscr{R}}_0$ defined in (\ref{strip R small_text}).
Again, the contribution of  $\Delta \neq 0$ to the powers of $R$ can be checked from the saturation time $t_s$ obtained in the field theories.
For a positive (negative) $\Delta$, in the small $R$ limit, the saturation time
$t_s$ becomes larger (smaller) than the case of $\Delta=0$  for  fixed $R$ and $T$.

Expanding  (\ref{t1_t2_t3_text}) for the small $\delta$ gives
$
t_1=\frac{y_b^{1+\frac{\Delta_0}{2}}\xi}{h(y_b)}+O(\delta^2),t_2=\boldsymbol{\mathscr{I}}(y_t)E+O(\delta^2)$, and $t_3=O(\delta^2) \,
$ where the function $\boldsymbol{\mathscr{I}}$ is defined by
\begin{equation}\label{I_scr_text}
\boldsymbol{\mathscr{I}}(y_b)\equiv\int_{0}^{y_b}\frac{y^{\Delta_0-\frac{\Delta}{2}}}{h(y)\sqrt{h(y) (\frac{y_{b}^{a_1(\tilde{d}-1)}}{y^{a_1(\tilde{d}-1)}}-1)}}\, .
\end{equation}
After collecting the results of $t_1$, $t_2$ and $t_3$ in their expansions for the small $\delta$ and the relation between $\delta$ and $\xi$ above, we arrive at
\be\label{late time time strip_text}
t-t_s= u_1\delta+O(\delta^2) \, ,
\ee
where
\be \label{u1} u_1=\frac{a_1}{4}(\tilde{d}-1)g(y_b)y_b^{\frac{\Delta}{2}-\frac{\Delta_0}{2}}\left[\frac{2y_b^{1+\Delta_0}}{a_1(\tilde{d}-1)h(y_b)^2\boldsymbol{\mathscr{R}}'(y_b)}-\boldsymbol{\mathscr{I}}(y_b)\right]\, . \ee
 For a continuous saturation, $t_s$ is great than $t$, so $u_1$ should be smaller than zero.
 Nevertheless, when $u_1>0$, it has been discussed in \cite{Liu-s} that $y_c$ is still far away from $y_t$ near the saturation so $y_t$ might jump to $y_b$ at $t=t_s$, causing a discontinuous saturation. In particular, in the AdS background, the entanglement entropy undergoes discontinuous saturation for both the large $R$ and small $R$ limits.
 In the case of the general anisotropic model, the constraints on the scaling parameters due to $u_1 <0$, giving the continuous saturation are
\be\label{large r conti condi_text}
\Big[a_1 (\tilde{d}-1)+2 \Delta_0 -3 \Delta -2\Big]<0 \,
\ee
for large $R$,
and
\be\label{small r conti condi_text}
\frac{4a_1(\tilde{d}-1)}{(2+\Delta)\tilde{\mathscr{I}}_0 \tilde{\mathscr{R}}_0}<1 \,
\ee
for small $R$
where $\tilde{\mathscr{I}}_0$ and $\tilde{\mathscr{R}}_0$ are defined in (\ref{stripR}) and (\ref{stripR1}) (see the details in Appendix).
Recall that the scaling parameters are constrained  by (\ref{bk_c}) from the black brane,  (\ref{nec1}),(\ref{nec2}), and (\ref{nec3}) due to null energy conditions, and (\ref{uv constraint strip}) and (\ref{delta_c}) from the solution of the extremal surface as well as
(\ref{delta'_c}) and (\ref{S_c_sphere}) by replacing   $d-1$ with $\tilde{d}-1$ for the sensible entanglement entropy. Also recall that the dimension $d \geq 3$ is considered in  the strip case.
 We will further analyze the criteria of the continuous saturation  in the Einstein-Axion-Dilaton theory later.
To deal with the integration of area, it is convenient to reformulate (\ref{area_strip}) as
\be\label{re_area_strip_text}
\frac{A_{\Gamma}}{A_{\partial\Sigma}}=\frac{A_{\Gamma v<0}}{A_{\partial\Sigma}}-\frac{A_{1}}{A_{\partial\Sigma}}+\frac{A_{2}}{A_{\partial\Sigma}}+\boldsymbol{\mathscr{A}}(y_t) \, ,
\ee
where
\be\label{a1a2_text}
A_{1}\equiv\int_{y_c}^{y_t}\frac{dy}{\sqrt{H(y)}},~~~A_{2}\equiv\int_{0}^{y_t}\left(\frac{1}{\sqrt{H(y)}}-\frac{1}{y^{\frac{a_1}{2}(\tilde{d}-1)-\frac{\Delta}{2}}\sqrt{h(y)}\sqrt{1-\frac{y^{a_1(\tilde{d}-1)}}{y_t^{a_1(\tilde{d}-1)}}}}\right)
\ee
and the function $\boldsymbol{\mathscr{A}}$ is defined by
\be\label{def_A_scr_text}
\boldsymbol{\mathscr{A}}(y_t)\equiv\int_{a_{UV}}^{y_t}\frac{1}{y^{\frac{a_1}{2}(\tilde{d}-1)-\frac{\Delta}{2}}\sqrt{h(y)}\sqrt{1-\frac{y^{a_1(\tilde{d}-1)}}{y_t^{a_1(\tilde{d}-1)}}}}\, .
\ee
Again, $\boldsymbol{\mathscr{A}}(y_b)=\frac{A_{\Gamma eq}}{A_{\partial\Sigma}}$.
Next we do the small $\delta$ expansion of $A_{\Gamma v<0}$ in (\ref{area_strip}) and $A_1$, $A_2$ in (\ref{a1a2_text}) where
$
\frac{A_{\Gamma v<0}}{A_{\partial\Sigma}}\simeq \frac{A_{1}}{A_{\partial\Sigma}}\simeq y_b^{\frac{\Delta}{2}+1-\frac{a_1}{2}(\tilde{d}-1)}\delta$, and $\frac{A_{2}}{A_{\partial\Sigma}}\simeq \frac{2y_t^{1+\frac{\Delta_0}{2}-\frac{a_1}{2}(\tilde{d}-1)}E}{a_1(\tilde{d}-1)h(y_t)} \, .
$
Note that the leading order term of $A_{1}$ is the same as $A_{\Gamma v<0}$ in both $h(y_b)\rightarrow 1$ (small $R$) or $h(y_b)\rightarrow 0$ (large $R$). In the end, we conclude that in both the large and small $R$ limits, the area reaches its saturated value in the way of
\be\label{strip late time area growth_text}
A_{\Gamma}-A_{\Gamma eq}=\frac{A_{\partial\Sigma}g(y_b)}{2h(y_b)}\left(\frac{\boldsymbol{\mathscr{A}}'(y_b)}{\boldsymbol{\mathscr{R}}'(y_b)}y_b^{\frac{a_1}{2}(\tilde{d}-1)}-1\right)y_b^{\frac{\Delta}{2}+1-\frac{a_1}{2}(\tilde{d}-1)}\delta+O(\delta^2)\propto (t-t_s)^2 \, .
\ee
Following \cite{Liu-s}, a straightforward calculation also shows that in the  large and small R limits $\frac{\boldsymbol{\mathscr{A}}'(y_b)}{\boldsymbol{\mathscr{R}}'(y_b)}y_b^{\frac{a_1}{2}(\tilde{d}-1)}$ is equal to  $1$, which leads to $A_{\Gamma} \rightarrow A_{\Gamma eq}$ quadratically in $t-t_s$.
However, in general $R$, the leading order terms of $A_{1}$ and $ A_{\Gamma v<0}$ in the small $\delta$ expansion will be different,
leading to the non-zero coefficient of the linear $\delta$ term in  (\ref{strip late time area growth_text})  where $A_{\Gamma}-A_{\Gamma eq}\propto t-t_s$. In general, the powers of  $t-t_s$ in $A_{\Gamma}-A_{\Gamma eq}$ are independent of
the scaling parameters as well as the spatial dimension \cite{Fonda}.  \\

\section{An example from the Einstein-Axion-Dilaton theory }\label{sec6}
Here we study the allowed scaling parameter regions given by all the constraints in the cases of the sphere and strip separately.
We give an explicit example by considering an anisotropic background in the Einstein-Axion-Dilaton theory  \cite{Dimitrio_1708}. The background metric obtained there is
  \be
  ds^2=a^2C_Re^{\phi(r)/2}r^{-\frac{2\theta}{dz}}\left(-r^2(f(r)dt^2+\sum_{i=2}^{d-1}dx_i^2)+C_zr^{2/z}dx_1^2+\frac{dr^2}{f(r)a^2r^2}\right)
  \ee
where $C_R$ and $C_z$ are constants. Also,  $f(r)=1-\left(\frac{r_h}{r}\right)^{d+(1-\theta)/z}\,$ and $e^{\phi(r)/2}=r^{\frac{\sqrt{\theta^2+3z(1-\theta)-3}}{\sqrt{6}z}}$. The temperature of the black brane is given by $T=\frac{|d+1-\theta|}{4\pi r_h}$. According to our notations, the scaling parameters read
 \bea
  &&a_y=-\frac{\sqrt{\theta^2+3z(1-\theta)-3}}{\sqrt{6}z}+\frac{2\theta}{dz}+2 \, ,\\
  &&a_0=\frac{\sqrt{\theta^2+3z(1-\theta)-3}}{\sqrt{6}z}-\frac{2\theta}{dz}+2 \, ,\\
 &&a_1=\frac{\sqrt{\theta^2+3z(1-\theta)-3}}{\sqrt{6}z}-\frac{2\theta}{dz}+\frac{2}{z} \, ,\\
 &&a_i=a_0 \quad\quad\quad\quad\quad\quad\quad\quad\quad\quad\quad\quad\quad\quad\quad \mbox{for}~i=2,3,.....d-1 \, .
 \eea
Thus, ${\Delta_h}=d+\frac{1-\theta}{z}$, $\Delta=\frac2z-2$, and $\Delta_0=0$. Since $\partial_v \Theta(v) =0$ for $v\neq 0$,
the null energy conditions in (\ref{nec1}), (\ref{nec2}), and (\ref{nec3}) reduce to two constraints
  \be\label{NEC example}
  \theta^2+3z(1-\theta)-3\geq0,~~(z-1)(1+3z-\theta)\geq 0 \,.
  \ee

In the sphere case with $a_1=a_2=...=a_{d-2}$,  the value of $z$ becomes $z=1$ giving  $\Delta_h=d+1-\theta$ and $\Delta=0$. While $z=1$, the null energy conditions (\ref{NEC example}) constrain the allowed regions of the parameter $\theta$ as
  \be\label{nec z=1}
  \theta\leq0~~~\mbox{and}~~~\theta\geq 3 \,.
  \ee
  {Also,  for $\Delta=0$, (\ref{h_v_c_sphere}) leads to $\Delta_h >3$ being consistent with the requirement of (\ref{bk_c}), giving}
  \be\label{constraint_spere2}
  d-2>\theta
  \ee
  {with which, the constraint (\ref{a1_c}) holds for  $d \geq 4$.  The constraint (\ref{delta'_c}) is satisfied  since $\Delta_0=0$ in this model.
  Together with (\ref{S_c_sphere}),
  \be\label{constraint_spere3}
  -\frac{2 \theta }{d}-\frac{d \sqrt{(\theta -3) \theta }}{\sqrt{6}}+\frac{\sqrt{(\theta -3) \theta }}{\sqrt{6}}+6>0 \, .
  \ee
We summarize the constraints on the scaling parameter $\theta$ in the Einstein-Axion-Dilaton theory for the sphere case by choosing different dimension $d$ in Fig.(1).

As for the strip case, collecting all constraints from (\ref{uv constraint strip}),(\ref{delta_c}),
(\ref{delta'_c}) and (\ref{S_c_sphere}) by replacing   $d-1$ with $\tilde{d}-1$ and combining them with the null energy conditions  (\ref{NEC example}) in this model gives the allowed parameter regions for $\theta$ and $z$ in dimension $d=3$ and $d=4$ as shown in Fig.(2). Also, due to (\ref{large r conti condi_text}) for large $R$ and (\ref{small r conti condi_text}) for small $R$, within the allowed parameter regions, the model undergoes discontinuous saturation as in the AdS background\cite{Liu-s} although  more general scaling parameters are involved.

\begin{figure}
\centering
\includegraphics[width=0.49\textwidth]{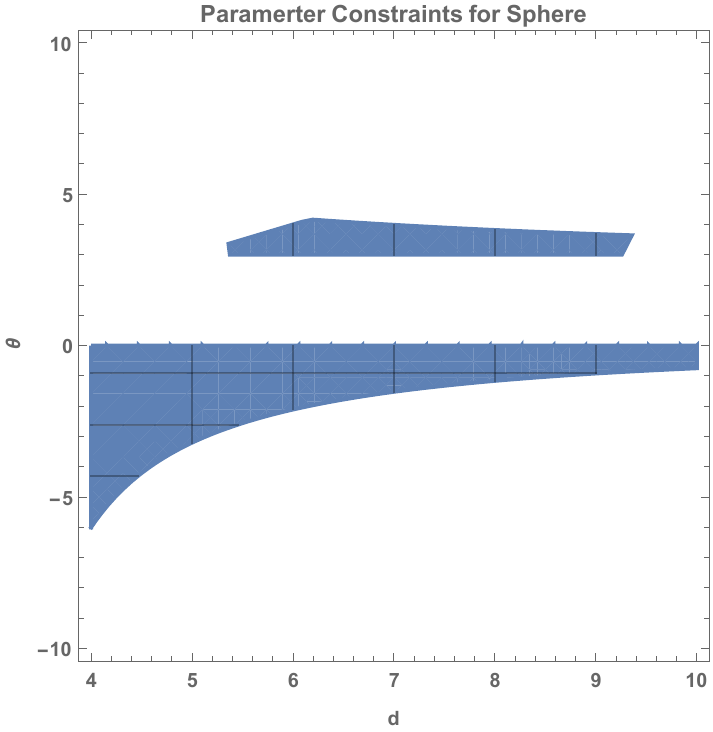}
\caption{{The allowed regions of $\theta$ by fixing $z=1$ in the sphere case in different dimension $d$, constrained by (\ref{nec z=1}), (\ref{constraint_spere2}) and (\ref{constraint_spere3}).}}
\end{figure}

\begin{figure}[h!]
\centering
\includegraphics[width=0.49\textwidth]{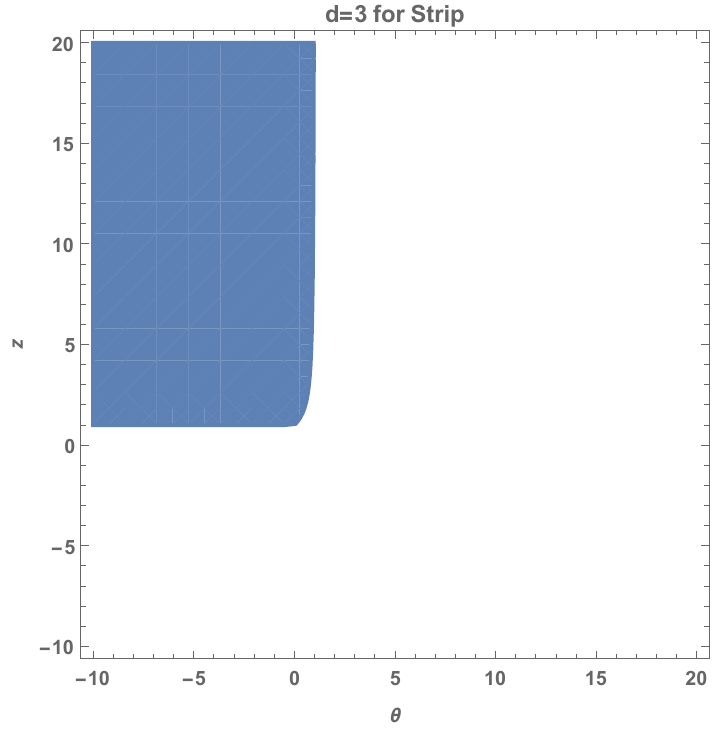}
\includegraphics[width=0.49\textwidth]{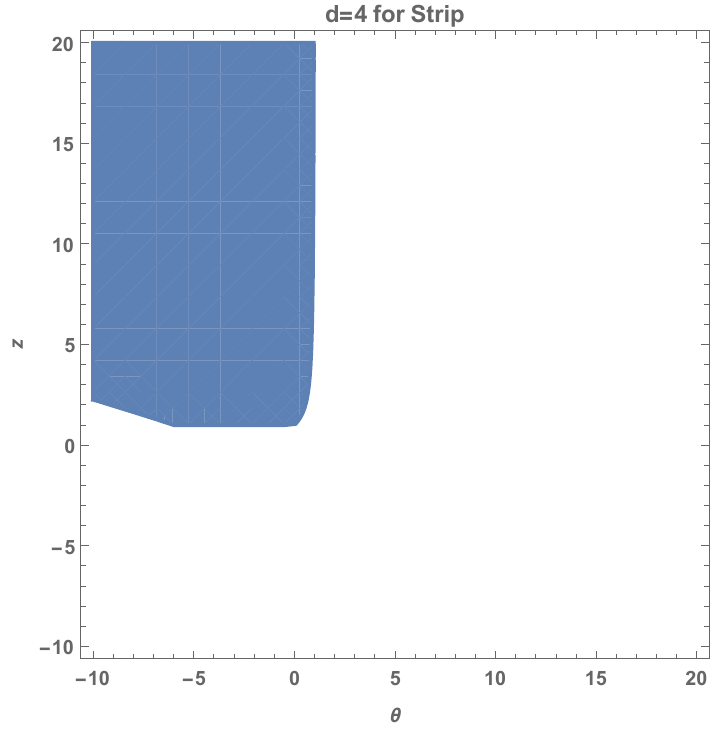}
\caption{{ The allowed regions of $\theta$ and $z$ in the Einstein-Axion-Dilaton theory in dimension $d=3$ (left figure) and $d=4$ (right figure), respectively constrained by NECs (\ref{NEC example}), the conditions of the existence of the solution for the strip case (\ref{uv constraint strip}) and (\ref{delta_c}), the condition of $\delta S(t)$ increasing in time for the early growth  (\ref{S_c_sphere}) via the replacement $d-1$ with $\tilde{d}-1$, and the thermodynamic stability condition (\ref{bk_c}). Note that the constraint in (\ref{delta'_c}) is satisfied since $\Delta_0=0$ in the Einstein-Axion-Dilaton theory. Together with (\ref{large r conti condi}) or (\ref{small r conti condi}), we find that the parameters in the blue regions always lead to discontinuous saturation for the system.}}
\end{figure}
\newpage

\section{Conclusions}\label{sec7}

In this paper, we employ the holographic method to study the thermalization of the strongly coupled Lifshitz-like anisotropic hyperscaling violation theories after a global quench. The gravity dual is the Vaydai-like geometry that describes the infalling of the massless delta function planar shell from the boundary and the subsequent formation of the black brane. We use the Ryu-Takayanagi formula to calculate the time evolution of the entanglement entropy between a strip of width $2R$ or a spherical region of radius $R$ and its outside region.
 Our model with the nonzero $\Delta=a_1+a_y-4$ generalizes the previous studies on the strongly coupled Lifshitz-like isotropic hyperscaling violation theories with $\Delta=0$.
 We find that quite generally the entanglement entropy grows polynomially in time with the power depending on the scaling parameters in both the early times and late times.
In particular, in the early time, for the positive (negative) value of $\Delta$, the power of $t$ increases (decreases) in $\Delta$  ($ |\Delta|$) to speed up (slow down) the growth rate as compared with the case of $\Delta=0$ for both the sphere and strip cases in the small and large $R$ limits.  In the late times,  as the boundary time $t$ reaches the saturation time scale $t_s$, the entanglement entropy is saturated in the same way as in the $\Delta=0$ case, which is
through the continuous saturation in the sphere case but the discontinuous saturation in the strip case in an example of an anisotropic background
in the Einstein-Axion-Dilaton theories. As for the saturation time $t_s$, by fixing the length scale $R$ and the temperature $T$, $t_s$ becomes larger (smaller) for a positive (negative) $\Delta$  than the case of $\Delta=0$ in both the strip and sphere cases in the large $R$ limit. For the sphere case in the small $R$ limit, in order to have an analytical expression of the saturation time $t_s$, one needs an exact solution of the extremal surface that can be found when $a_y=2$. Thus, in this case,  since $a_1$  is always positive resulting from the above mentioned constraints, the saturation time $t_s$ becomes larger (smaller) as $a_1$ increases
(decreases). For the strip case in the small $R$ limit, for a positive (negative) $\Delta$ the saturation time scale
$t_s$ becomes larger (smaller) than the case of $\Delta=0$, again for the fixed $R$ and $T$.
These behaviors can in principle be tested experimentally and compared with other methods to characterize the thermalization of  nonequilibrium systems.

\begin{appendices}

\section{The Criteria of Continuous/Discontinuous Saturation for the strip case}\label{app}
In section (\ref{EE in strip late time_text}), we assume that $y_c$ is close to $y_t$ near the saturation, and derive (\ref{late time time strip_text}). However, in the case of $u_1>0$,  $y_c$ is still far away from $y_b$ near the saturation, and $y_t$ will jump to $y_b$ at $t=t_s$. In this appendix, we will find  the condition for $u_1<0$ in the large and small $R$ limits. The functions, $\boldsymbol{\mathscr{R}}(y_t)$ and $\boldsymbol{\mathscr{I}}(y_b)$ in (\ref{r_scr_text}) and (\ref{I_scr_text}) can be rewritten in the summation forms
\bea\label{stripR}
 \boldsymbol{\mathscr{R}}(y_t)&&\equiv \int_{0}^{y_t}\frac{y^{\frac{a_1}{2}(\tilde{d}-1)+\frac{\Delta}{2}}}{y_t^{\frac{a_1}{2}(\tilde{d}-1)}\sqrt{h(y)}\sqrt{1-\frac{y^{a_1(\tilde{d}-1)}}{y_t^{a_1(\tilde{d}-1)}}}}\,dy\nonumber\\
 &&=\frac{y_t^{\frac{\Delta}{2}+1}}{a_1(\tilde{d}-1)}\sum _{j=0}^{\infty } \tilde{\mathscr{R}}_j\beta^{j \Delta_h}, ~~~\tilde{\mathscr{R}}_j=\frac{\Gamma \left(j+\frac{1}{2}\right) \Gamma\left(\frac{1}{2}+\frac{\Delta+2}{2a_1(\tilde{d}-1)}+\frac{j\Delta_h}{a_1(\tilde{d}-1)}\right)}{\Gamma (j+1)\Gamma\left(1+\frac{\Delta+2}{2a_1(\tilde{d}-1)}+\frac{j\Delta_h}{a_1(\tilde{d}-1)}\right)}\, ,\label{strip R sries}
\eea
\bea \label{stripR1}
\boldsymbol{\mathscr{I}}(y_b)&&\equiv\int_{0}^{y_b}\frac{y^{\Delta_0-\frac{\Delta}{2}}}{h(y)\sqrt{h(y) (\frac{y_{b}^{a_1(\tilde{d}-1)}}{y^{a_1(\tilde{d}-1)}}-1)}}\nonumber\\
&&=\frac{2y_b^{1+\Delta_0-\frac{\Delta}{2}}}{a_1(\tilde{d}-1)}\sum_{j=0}^{\infty}\tilde{\mathscr{I}}_j\beta^{j\Delta_h},~~~\tilde{\mathscr{I}}_j\equiv\frac{\Gamma\left(j+\frac{3}{2}\right)\Gamma\left(\frac{2\Delta_h+\Delta_0-\Delta}{2a_1(\tilde{d}-1)}+\frac{j\Delta_h}{a_1(\tilde{d}-1)}\right)}{\Gamma\left(j+1\right)\Gamma\left(\frac{1}{2}+\frac{2\Delta_h+\Delta_0-\Delta}{2a_1(\tilde{d}-1)}+\frac{j\Delta_h}{a_1(\tilde{d}-1)}\right)}\label{Iexp}
\, , \eea
where $\beta=\frac{y_b}{y_h}$. In (\ref{stripR}) and (\ref{Iexp}), the binomial identity and the Euler integral of the first kind have been applied.  From (\ref{strip R sries}) and $\boldsymbol{\mathscr{R}}(y_b)=R$, it is straightforward to obtain
\be\label{r'}
\boldsymbol{\mathscr{R}}'(y_b)=\left(\frac{\Delta}{2}+1\right)\frac{R}{y_b}
+\frac{y_b^{\frac{\Delta}{2}}\Delta_h}{a_1(\tilde{d}-1)}\sum_{j=0}^{\infty}j\tilde{\mathscr{R}}_j\beta^{j\Delta_h} \, .\ee
In the large $R$ limit, the tip of extremal surface $y_b$ is very close to the horizon $y_h$ for $\beta\rightarrow 1$. In the small $R$ limit, the tip of the extremal surface instead is very close to the boundary $y=0$ for $\beta\rightarrow 0$.

The leading order behaviour of $\boldsymbol{\mathscr{I}}(y_b)$ in the large $R$ limit relies on the asymptotic approximation  in (\ref{Iexp}) at $j\rightarrow \infty$ given by
\be\label{large i}
\tilde{\mathscr{I}}_j\beta^{j\Delta_h}\simeq \sqrt{\frac{a_1(\tilde{d}-1)}{\Delta_h}}\beta^{\Delta_h j}-\frac{\sqrt{a_1(\tilde{d}-1)}}{8\Delta_h^{3/2}j}\left(\Delta_h+2\Delta_0-2\Delta-a_1(\tilde{d}-1)\right)\beta^{\Delta_h j} \, .
\ee
Then in the large $R$ limit (\ref{Iexp}) can be approximated by
\begin{align}\label{196}
    \boldsymbol{\mathscr{I}}(y_b)&\simeq \frac{2y_b^{1+\Delta_0-\frac{\Delta}{2}}}{a_1(\tilde{d}-1)}\sum_{j=1}^{\infty}\left(\sqrt{\frac{a_1(\tilde{d}-1)}{\Delta_h}}\beta^{\Delta_h j}-\frac{\sqrt{a_1(\tilde{d}-1)}}{8\Delta_h^{3/2}j}\left(\Delta_h+2\Delta_0-2\Delta-a_1(\tilde{d}-1)\right)\beta^{\Delta_h j}\right)\\
&\simeq \frac{y_h^{\Delta_0-\Delta}}{\Delta_h \gamma}\left[\frac{1}{\epsilon}+\frac{\ln(\epsilon)}{8}\left(\Delta_h+2\Delta_0-2\Delta-a_1(\tilde{d}-1)\right)\right]\label{197}\\
&\simeq \frac{y_h^{\Delta_0-\Delta}}{\Delta_h \gamma}e^{2\gamma R}-\frac{y_h^{\Delta_0-\Delta}}{4 \Delta_h}\left(\Delta_h+2\Delta_0-2\Delta-a_1(\tilde{d}-1)\right)R \, .\label{large R I R}
\end{align}
Apparently when $\beta\rightarrow 1$, the summation above diverges. This divergence can be translated into the singular behavior of $ \boldsymbol{\mathscr{I}}(y_b)$
 in the case of $y_b= y_h(1-\epsilon)$, when $\epsilon \rightarrow 0$, whereas  the most singular behaviour is given by (\ref{197}). Moreover, the last expression is obtained using the relation $\ln(\epsilon)= -2R\gamma+O(R^0)$. Similarly, to discover how $\boldsymbol{\mathscr{R}}'(y_b)$ behaves in the large $R$ limit, the associated large $j$ behaviour  in (\ref{r'}) is obtained as
\be j\tilde{\mathscr{R}}_j\beta^{j\Delta_h}\simeq \sqrt{\frac{a_1(\tilde{d}-1)}{\Delta_h}}\beta^{\Delta_h j}-\frac{\sqrt{a_1(\tilde{d}-1)}}{8\Delta_h^{3/2}j}\left(4+\Delta_h+2\Delta+a_1(\tilde{d}-1)\right)\beta^{\Delta_h j}\, .\ee
Then  the leading term when $\beta\rightarrow 1$ becomes
\begin{align}\label{large strip r'}
\boldsymbol{\mathscr{R}}'(y_b)\simeq &\left(\frac{\Delta}{2}+1\right)\frac{R}{y_b}\nonumber\\
&+\frac{y_b^{\frac{\Delta}{2}}\Delta_h}{a_1(\tilde{d}-1)}\sum_{j=1}^{\infty}\left(\sqrt{\frac{a_1(\tilde{d}-1)}{\Delta_h}}\beta^{\Delta_h j}-\frac{\sqrt{a_1(\tilde{d}-1)}}{8\Delta_h^{3/2}j}\left(4+\Delta_h+2\Delta+a_1(\tilde{d}-1)\right)\beta^{\Delta_h j}\right)\\
&\simeq \frac{1}{2 y_h \gamma}\left[\frac{1}{\epsilon}+\frac{\ln(\epsilon)}{8}\left(4+\Delta_h+2\Delta+a_1(\tilde{d}-1)\right)\right]+\left(\frac{\Delta}{2}+1\right)\frac{R}{y_h}\label{197r}\\
&\simeq \frac{e^{2\gamma R}}{2 y_h \gamma}+
\left[\frac{\Delta}{2}+1-
\frac{\left(4+\Delta_h+2\Delta+a_1(\tilde{d}-1)\right)}{8}\right]\frac{R}{y_h}\label{large R R R}
\end{align}
As a result, due to (\ref{large R I R}) and (\ref{large R R R}), the behavior of $u_1$ (\ref{u1}) in the large $R$ limit
is
\be\label{u1 large R}
u_1\simeq \frac{a_1(\tilde{d}-1)}{16 \Delta_h}y_h^{\frac{\Delta_0}{2}-\frac{\Delta}{2}}\Big[a_1 (\tilde{d}-1)+2 \Delta _0-3 \Delta -2\Big]R
\ee
Note that we have used $h(y_b)\simeq \Delta_h e^{-2\gamma R}$. Since $\Delta_h>0$ is required,  from (\ref{uv constraint strip}) with $a_1>0$, we find that
\be\label{large r conti condi}
u_1<0\Leftrightarrow \Big[a_1 (\tilde{d}-1)+2 \Delta _0-3 \Delta -2\Big]<0
\ee
for the large $R$ limit.
In the small $R$ limit, we can approximate (\ref{Iexp}) and (\ref{r'}) as
\be\label{small R I}
\boldsymbol{\mathscr{I}}(y_b)\simeq \frac{y_b^{1+\Delta_0-\frac{\Delta}{2}}}{a_1(\tilde{d}-1)}\tilde{\mathscr{I}}_0
\ee
and
\be\label{small R R'}
\boldsymbol{\mathscr{R}}'(y_b)\simeq \frac{\Delta+2}{2a_1(\tilde{d}-1)}y_b^{\frac{\Delta}{2}}\tilde{\mathscr{R}}_0 \, .
\ee
Then we  find  (\ref{u1}) in the  small $R$ limit to be
\be\label{u1 small R}
u_1\simeq \frac{g(y_b)}{4}\left[\frac{4a_1(\tilde{d}-1)}{(2+\Delta)\tilde{\mathscr{I}}_0\tilde{\mathscr{R}}_0}-1\right]\tilde{\mathscr{I}}_0 y_b^{1+\frac{\Delta_0}{2}} \,.
\ee
By (\ref{uv constraint strip}), (\ref{Iexp}) and (\ref{r'}) where $\tilde{\mathscr{I}}_0>0$, $\tilde{\mathscr{R}}_0>0$ and $a_1>0$, we then arrive at $\frac{4a_1(\tilde{d}-1)}{(2+\Delta)\tilde{\mathscr{I}}_0\tilde{\mathscr{R}}_0}>0$. Finally, from (\ref{u1 small R}), we conclude
\be\label{small r conti condi}
u_1<0\Leftrightarrow \frac{4a_1(\tilde{d}-1)}{(2+\Delta)\tilde{\mathscr{I}}_0\tilde{\mathscr{R}}_0}<1
\ee
for the small $R$ limit.
\end{appendices}

\begin{acknowledgments}
 This work was supported in part by the
Ministry of Science and Technology, Taiwan.
\end{acknowledgments}

\end{document}